\documentclass[a4paper, 12pt]{article}

\usepackage{lineno,hyperref}
\DeclareMathAlphabet{\pazocal}{OMS}{zplm}{m}{n}

\usepackage{dsfont}
\usepackage{amsmath}
\usepackage{subfig}
\usepackage{float}
\usepackage{longtable}
\usepackage{booktabs}
\usepackage{lscape}
\usepackage[flushleft]{threeparttable}
\usepackage{multirow}
\usepackage{afterpage}
\usepackage{hyperref}
\usepackage{natbib}

\modulolinenumbers[5]

\RequirePackage{fix-cm}
\usepackage{latexsym}
\usepackage{lineno,hyperref}
\usepackage{graphics}
\usepackage{graphicx}
\usepackage{amsmath}
\usepackage{amsfonts}
\usepackage{amssymb}
\usepackage[dvipsnames]{xcolor}
\usepackage{tikz}
\usepackage{color}
\usepackage{xcolor}
\usepackage{natbib}
\usepackage{array}
\usepackage{subfig}
\usepackage{caption}
\usepackage{multicol}
\usepackage{multirow}
\usepackage{acro}
\usepackage{relsize}
\usepackage[ruled,vlined]{algorithm2e}
\usepackage{footnote}
\usepackage{lipsum} 
\usepackage{makecell}

\def\argmin{\mathop{\rm arg\,min}\limits}

\makeatletter
\newcommand{\thickhline}{%
    \noalign {\ifnum 0=`}\fi \hrule height 1pt
      \futurelet \reserved@a \@xhline
}

\newcolumntype{'}{@{\hskip\tabcolsep\vrule width 1pt\hskip\tabcolsep}}
\makeatother

\begin{document}

\title{Comparing clusterings and numbers of clusters by aggregation of calibrated clustering validity indexes}

\author{Serhat Emre Akhanli (1) and Christian Hennig (2)\\
\small
(1) Department of Statistical Science,\\ \small University College London, UK\\ 
\small and Department of Statistics, Faculty of Science,\\
\small Mu\u{g}la S{\i}tk{\i} Ko\c{c}man University, Mu\u{g}la, Turkey \\
              \small Tel.: +905459267047\\
\small (2) Dipartimento di Scienze Statistiche ``Paolo Fortunati''\\
\small        Universita di Bologna\\
\small        Bologna, Via delle belle Arti, 41, 40126, Italy
 }


\maketitle

\begin{abstract}
A key issue in cluster analysis is the choice of an appropriate clustering method and the determination of the best number of clusters. Different clusterings are optimal on the same data set according to different criteria, and the choice of such criteria depends on the context and aim of clustering. Therefore, researchers need to consider what data analytic characteristics the clusters they are aiming at are supposed to have, among others within-cluster homogeneity, between-clusters separation, and stability. Here, a set of internal clustering validity indexes measuring different aspects of clustering quality is proposed, including some indexes from the literature. Users can choose the indexes that are relevant in the application at hand. In order to measure the overall quality of a clustering (for comparing clusterings from different methods and/or different numbers of clusters), the index values are calibrated for aggregation. Calibration is relative to a set of random clusterings on the same data. Two specific aggregated indexes are proposed and compared with existing indexes on simulated and real data.
\end{abstract}

\section{Introduction}
{\it This version has been accepted by Statistics and Computing for publication.}

Cluster validation, which is the evaluation of the quality of clusterings, is crucial in cluster analysis in order to make sure that a given clustering makes sense, but also in order to compare different clusterings. These may stem from different clustering methods, but may also have different numbers of clusters, and in fact optimizing measurements of clustering validity is a main approach to estimating the number of clusters, see \cite{hennig2015clustering26}.

In much literature on cluster analysis it is assumed, implicitly or explicitly, that there is only a single ``true'' clustering for a given data set, and that the aim of cluster analysis is to find that clustering. According to this logic, clusterings are better or worse depending on how close they are to the ``true'' clustering. If a true grouping is known (which does not necessarily have to be unique), ``external'' cluster validation compares a clustering to the true clustering (or more generally to existing external information). A popular external clustering validity index is the Adjusted Rand Index (\cite{HubertArabie1985}). Here we are concerned with ``internal'' cluster validation (sometimes referred to as ``relative cluster validation'' when used to compare different clusterings, see \cite{jaindubes}), evaluating the cluster quality without reference to an external ``truth'', which in most applications of course is not known.

\cite{hennig2015clustering31,hennig15true} argued that the ``best'' clustering depends on background information and the aim of clustering, and that different clusterings can be optimal in different relevant respects on the same data. As an example, within-cluster homogeneity and between-clusters separation are often mentioned as major aims of clustering, but these two aims can be conflicting. There may be widespread groups of points in the data without ``gaps'', which can therefore not be split up into separated subgroups, but may however contain very large within-group distances. If clustering is done for shape recognition, separation is most important and such groups should not be split up. If clustering is used for database organisation, for example allowing to find a set of very similar images for a given image, homogeneity is most important, and a data subset containing too large distances needs to be split up into two or more clusters. Also other aspects may matter such as approximation of the dissimilarity relations in the data by the clustering structure, or distributional shapes (e.g., linearity or normality). Therefore the data cannot decide about the ``optimal'' clustering on their own, and user input is needed in any case. 

Many existing clustering validity indexes attempt to measure the quality of a clustering by a single number, see \cite{hennig2015clustering26} for a review of such indexes. Such indexes are sometimes called ``objective'', because they do not require decisions or tuning by the users. This is certainly popular among users who do not want to make such decisions, be it for lack of understanding of the implications, or be it for the desire to ``let the data speak on their own''. But in any case such users will need to decide which index to trust for what reason, and given that requirements for a good clustering depend on the application at hand, it makes sense that there is a choice between various criteria and approaches. But the literature is rarely explicit about this and tends to suggest that the clustering problem can and should be solved without critical user input.  

There is a tension in statistics between the idea that analyses should be very closely adapted to the specifics of the situation, necessarily strongly involving the researchers' perspective, and the idea that analyses should be as ``objective'' and independent of a personal point of view as possible, see \cite{gelmanhennig17}. A heavy focus on user input will give the user optimal flexibility to take into account background knowledge and the aim of clustering, but the user may not feel able to make all the required choices, some of which may be very subtle and may be connected at best very indirectly to the available information. Furthermore, it is hard to systematically investigate the quality and reliability of such an approach, because every situation is different and it may be unclear how to generalise from one situation to another. On the other hand, a heavy focus on ``objective'' unified criteria and evaluation over many situations will make it hard or even impossible to do the individual circumstances justice. In the present paper we try to balance these two aspects by presenting a framework that allows for very flexible customisation, while at the same time proposing two specific aggregated indexes as possible starter tools for a good number of situations that allow us to systematically evaluate the approach on simulated and benchmark data.
  
Many validity indexes balance a small within cluster heterogeneity and a large between-clusters heterogeneity in a certain way, such as Average Silhouette Width (ASW; \cite{kaufman2009finding}) or the Calinski and Harabasz index (CH; \cite{calinski1974dendrite}),  whereas others have different goals; for example Hubert's $\Gamma$ index (\cite{hubert1976quadratic}) emphasises good representation of the dissimilarity structure by the clustering. In most applications, various desirable characteristics need to be balanced against each other. It is clear that it is easier to achieve homogeneous clusters if the number of clusters is high, and better cluster separation if the number of clusters is low, but in different applications these objectives may be weighted differently, which cannot be expressed by a single index.

The approach taken here, first introduced in \cite{hennig2017cluster}, is to consider a collection of validity indexes that measure various aspects of cluster quality in order to allow the user to weight and aggregate them to a quality measurement adapted to their specific clustering aim. This can then be used to decide between different clusterings with different numbers of clusters or also from different clustering methods. Particular attention is paid to the issue of making the values of the different indexes comparable when aggregating them, allowing for an interpretation of weights in terms of relative importance. This is done by generating random clusterings over the given data set and by using the distribution of the resulting index values for calibration.

Some authors have already become aware of the benefits of looking at several criteria for comparing clusterings, and there is some related work on multi-objective clustering, mostly about finding the set of Pareto optimal solutions, see, e.g., \cite{delattrehansen80} and the overview in \cite{handlknowles15}.

Section~\ref{sec:general_notation} introduces the notation. Section~\ref{sec:internal_validation_indexes} is devoted to clustering validity indexes. It has three subsections introducing clustering validity indexes from the literature, indexes measuring specific aspects of cluster validity to be used for aggregation, and resampling-based indexes measuring cluster stability. Section~\ref{sec:aggregation_indexes} describes how an aggregated index can be defined from several indexes measuring specific characteristics, including calibration by random clusterings. Section~\ref{sec:numerical_experiment} proposes two specific aggregated indexes for somewhat general purposes, presents a simulation study comparing these to indexes from the literature, and uses these indexes to analyse three real data sets with and one without given classes. 
Section~\ref{sec:discussion} concludes the paper.

\section{General notation}
\label{sec:general_notation}

Given a data set, i.e., a set of distinguishable objects $\mathcal{X}=\left\{ x_{1}, x_{2}, \ldots, x_{n} \right\}$, the aim of cluster analysis is to group them into subsets of $\mathcal{X}$. A clustering is denoted by $\mathcal{C}=\left\{ C_{1}, C_{2}, \ldots, C_{K} \right\}$, $C_k\subseteq \mathcal{X}$ with cluster size $n_{k}= |C_{k}|,\ k=1,\ldots,K$. We require $\mathcal{C}$ to be a partition, e.g., $k \neq g  \Rightarrow C_{k} \cap C_{g} = \emptyset$ and $\bigcup_{k=1}^{K} C_{k} = \mathcal{X}$. Clusters are assumed to be crisp rather than fuzzy, i.e., an object is either a full member of a cluster or not a member of this cluster at all.
An alternative way to write $x_{i}\in C_k$ is $l_i=k$, i.e., $l_i\in\{1,\ldots,K\}$ is the cluster label of $x_{i}$.

The approach presented here is defined for general dissimilarity data. A dissimilarity is a function $d : \mathcal{X}^{2} \rightarrow \mathbb{R}_{0}^{+}$ so that $d(x_{i}, x_{j}) = d(x_{j}, x_{i}) \geq 0$ and $d(x_{i}, x_{i}) = 0$ for $x_{i}, x_{j} \in \mathcal{X}$. Many dissimilarities are distances, i.e., they also fulfill the triangle inequality, but this is not necessarily required here. Dissimilarities are extremely flexible. They can be defined for all kinds of data, such as functions, time series, categorical data, image data, text data etc. If data are Euclidean, obviously the Euclidean distance can be used, which is what will be done in the later experiments. See \citet{hennig2015clustering31} for a more general overview of dissimilarity measures used in cluster analysis.

\section{Clustering Validity Measurement}
\label{sec:internal_validation_indexes}

This section lists various measurements of clustering validity. It has three parts. In the first part we review some popular indexes that are proposed in the literature. Each of these indexes was proposed with the ambition to measure clustering quality on its own, so that a uniquely optimal clustering or number of clusters can be found by optimizing one index. In the second part we list indexes that can be used to measure a single isolated aspect of clustering validity with a view of defining a composite measurement, adapted to the requirements of a specific application, by aggregating several of these indexes. In the third part we review resampling-based measurements of clustering stability. 

\cite{hennig2017cluster} suggested to transform indexes to be aggregated into the $[0,1]$-range so that for all indexes bigger values mean a better clustering quality. However, as acknowledged by \cite{hennig2017cluster}, this is not enough for making index values comparable, and here we give the untransformed forms of the indexes.

All these indexes are internal, i.e., they can be computed for a given partition $\mathcal{C}$ on a data set $\mathcal{X}$, often equipped with a dissimilarity $d$.  The indexes are either not defined or take trivial values for $K=1$, so using them for finding an optimal number of clusters assumes $K\ge 2$.

\subsection{Some popular clustering quality indexes}
\label{subsec:popular_cquality}

Here are some of the most popular of the considerable number of clustering quality indexes that have been published in the literature. All of these were meant for use on their own, although they may in principle also be used as part of a composite index. But most of these indexes attempt to balance two or more aspects of clustering quality, and from our point of view, for defining a composite index, it is preferable to use indexes that measure different aspects separately (as introduced in Section \ref{subsec:aspect_cquality}), because this improves the clarity of interpretation of the composite index. Unless indicated otherwise, for these indexes a better clustering is indicated by a larger value, and the best number of clusters can be chosen by maximizing any of them over $K$, i.e., comparing solutions from the same clustering method with different fixed values of $K$. 
 
\begin{description}
	\item[\textbf{The Average Silhouette Width (ASW)}] \cite{kaufman2009finding} compare the average dissimilarity of an observation to members of its own cluster to the average dissimilarity to members of the closest cluster to which it is not classified. It was one of the best performers for estimating the number of clusters in the comparative study of \cite{AGMPP12}.
For $i = 1, \ldots, n$, define the ``silhouette width'' 
		\begin{displaymath}
			s_{i}=\frac{b_{i}-a_{i}}{\max{\left\{a_{i}, b_{i}\right\}}} \in [-1,1],
		\end{displaymath}
where		
		\begin{align*}
			a_{i}=\frac{1}{n_{l_i}-1} \sum_{x_{j} \in C_{l_i}} d(x_{i}, x_{j}),\\
			b_{i}=\min_{h \neq l_i} \frac{1}{n_{h}} \sum_{x_{j} \in C_{h}} d(x_{i}, x_{j}). 
		\end{align*}
The ASW is then defined as
	\begin{displaymath}
		I_{ASW}(\mathcal{C}) = \frac{1}{n} \sum_{i=1}^{n} s_{i}.
	\end{displaymath}
%
%
		
	\item[\textbf{The Calinski-Harabasz index (CH)}] \cite{calinski1974dendrite}: This index compares squared within-cluster dissimilarities (measuring homogeneity) with squared dissimilarities between cluster means (measuring separation). This was originally defined for Euclidean data and use with $K$-means (the form given here is equal to the original form with $d$ as Euclidean distance).
It achieved very good results in the comparative study by \cite{milligancooper}. It is defined as
		\begin{displaymath}
			I_{CH}(\mathcal{C}) = \frac{\mathbf{B}(n-K)}{\mathbf{W}(K-1)},
		\end{displaymath}
where
		
		\begin{align*}
			\mathbf{W} &=\sum_{k=1}^{K} \frac{1}{n_{k}} \sum_{x_{i}, x_{j} \in C_{k}} d(x_{i}, x_{j})^{2},\\
			\mathbf{B} &=\frac{1}{n} \sum_{i,j=1}^{n} d(x_{i}, x_{j})^{2} - \mathbf{W}. 
		\end{align*}	
\item[\textbf{The Dunn  Index}] (\cite{dunn1974well}) compares the minimum distance between any two clusters with the maximum distance within a cluster:

	\begin{align*}
		I_{Dunn}(\mathcal{C}) &= \frac{\min_{1 \leq g < h \leq K}\min_{x_{i} \in C_{g}, x_{j} \in C_{h}} d(x_{i}, x_{j})}{\max_{1 \leq k \leq K}  \max_{x_{i},x_{j} \in C_k} d(x_{i},x_{j})} \\
		&\in [0,1].
	\end{align*}
	
	\item[\textbf{Clustering Validity Index Based on Nearest}] \textbf{Neighbours (CVNN)} was proposed by \citet{liu2013understanding} for fulfilling a number of desirable properties. Its separation statistic is based on local neighbourhoods of the points in the least separated cluster, looking at their $\kappa$ nearest neighbours:
\begin{displaymath}
			I_{Sep}(\mathcal{C}; \kappa) = \max_{1 \leq k \leq K} \left( \frac{1}{n_{k}} \sum_{x \in C_{k}} \frac{q_{\kappa}(x)}{\kappa} \right),
\end{displaymath}
where $q_{\kappa}(x)$ is the number of observations among the $\kappa$ (to be fixed by the user) nearest neighbours of $x$ that are not in the same cluster. A compactness statistics $I_{Com}(\mathcal{C})$ is just the average of all within-cluster dissimilarities. The CVNN index aggregates these:
	\begin{displaymath}
		I_{CVNN}(\mathcal{C}, \kappa) = \frac{I_{Sep}(\mathcal{C}, \kappa)}{\max_{\mathcal{C} \in \mathcal{K}} I_{Sep}(\mathcal{C}, \kappa) } + \frac{I_{Com}(\mathcal{C})}{\max_{\mathcal{C} \in \mathcal{K}} I_{Com}(\mathcal{C}) },
	\end{displaymath}	
where $\mathcal{K}$ is the set of all considered clusterings. Here smaller values indicate a better clustering; CVNN needs to be minimised in order to find an optimal $K$.

	\item[\textbf{Pearson$\Gamma$ (PG):}] \cite{hubert1976quadratic} introduced a family of indexes called (Hubert's) $\Gamma$ measuring the quality of fit of a dissimilarity matrix by some representation, which could be a clustering. More than one version of $\Gamma$ can be used for clustering validation; the simplest one is based on the Pearson sample correlation $\rho$. It interprets the ``clustering induced dissimilarity'' $\mathbf{c} = vec([c_{ij}]_{i<j})$, where $c_{ij} = \mathbf{1}(l_{i} \neq l_{j})$, i.e. the indicator whether $x_i$ and $x_j$ are in different clusters, as a ``fit'' of the given data dissimilarity 
$\mathbf{d} = vec\left([d(x_{i}, x_{j})]_{i<j}\right)$,
and measures its quality as
	\begin{displaymath}
		I_{Pearson \Gamma}(\mathcal{C}) = \rho(\mathbf{d}, \mathbf{c}).
	\end{displaymath}
This index can be used on its own to measure clustering quality. It can also be used as part of a composite index, measuring a specific aspect of clustering quality, namely the approximation of the dissimilarity structure by the clustering.
In some applications clusterings are computed to summarise dissimilarity information, potentially for use of the cluster indicator as explanatory factor in an analysis of variance or similar, in which case the representation of the dissimilarity information is the central clustering aim. 
\end{description}

\subsection{Measurement of isolated aspects of clustering quality}
\label{subsec:aspect_cquality}
The following indexes measure isolated aspects of clustering quality. They can be used to compare different clusterings, but when used for comparing different numbers of clusters, some of them will systematically prefer either a smaller or larger number of clusters when used on their own. For example, it is easier to achieve smaller average or maximum within-cluster distances with a larger number of smaller clusters. So these indexes will normally be used as part of a composite index when deciding the number of clusters.

\begin{description}
	\item[\textbf{Average within-cluster dissimilarities:}] Most informal descriptions of what a ``cluster'' is involve homogeneity in the sense of high similarity or low dissimilarity of the objects within a cluster, and this is relevant in most applications of clustering. There are various ways of measuring whether within-cluster dissimilarities are generally low. A straightforward index averages all within-cluster dissimilarities in such a way that every observation has the same overall weight. Alternatives could for example involve squared distances or look at the maximum within-cluster distance. 

	\begin{displaymath}
		I_{ave.wit}(\mathcal{C}) = \frac{1}{n} \sum_{k=1}^{K} \frac{1}{n_k-1}\sum_{x_{i} \neq x_{j} \in C_{k}} d(x_{i},x_{j}). 
	\end{displaymath}	
A smaller value indicates better clustering quality. 


\item[\textbf{Separation index:}] Most informal descriptions of what makes a cluster mention between-cluster separation besides within-cluster homogeneity. Separation measurement should optimally focus on objects on the ``border'' of clusters. It would be possible to consider the minimum between-clusters dissimilarity (as done by the Dunn index), but this might be inappropriate, because in the case of there being more than two clusters the computation only depends on the two closest clusters, and represents a cluster only by a single point, which may be atypical. On the other hand, looking at the distance between cluster means as done by the CH index is not very informative about what goes on ``between'' the clusters. Thus, we propose another index that takes into account a portion, $p$, of objects in each cluster that are closest to another cluster. 
	
	For every object $x_{i} \in C_{k}$, $i = 1, \ldots, n$, $k \in {1, \ldots, K}$, let $d_{k:i} = \min_{x_{j} \notin C_{k}} d(x_{i},x_{j})$. Let $d_{k:(1)} \leq \ldots \leq d_{k:(n_{k})}$ be the values of $d_{k:i}$ for $x_{i} \in C_{k}$ ordered from the smallest to the largest, and let $[pn_{k}]$  be the largest integer $\leq pn_{k}$. Then, the separation index with the parameter $p$ is defined as
	\begin{displaymath}
		I_{sep.index}(\mathcal{C};p) = \frac{1}{\sum_{k=1}^{K} [pn_{k}]} \sum_{k=1}^{K} \sum_{i=1}^{[pn_{k}]} d_{k:(i)},  
	\end{displaymath}
	
Larger values are better. The proportion $p$ is a tuning parameter specifying what percentage of points should contribute to the ``cluster border''. We suggest $p=0.1$ as default.  

\item[\textbf{Widest within-cluster gap:}] This index measures within-cluster homogeneity in a quite different way, considering the biggest dissimilarity $d_g$ so that the cluster could be split into two subclusters with all dissimilarities between these subclusters $\ge d_g$. This is relevant in applications in which good within-cluster connectivity is required, e.g., in the delimitation of biological species using genetic data; species should be genetically connected, and a gap between subclusters could mean that no genetic exchange happens between the subclusters (on top of this, genetic separation is also important).

	\begin{equation}
		I_{widest.gap}(\mathcal{C}) = \max_{C \in \mathcal{C}, \; D,E: \; C = D \cup E} \quad \min_{x_{i} \in D , \; x_{j} \in E} d(x_{i},x_{j}).  
		\label{eq:widest_gap}
	\end{equation}
Smaller values are better.	

\item[\textbf{Representation of dissimilarity structure by}]
\textbf{clustering:} Clusterings are used in some applications to represent the more complex information in the full dissimilarity matrix in a simpler way, and it is of interest to measure the quality of representation in some way. For this aim here we use Pearson$\Gamma$ as defined above.   

\item[\textbf{Uniformity of cluster sizes:}]
Although not normally listed as primary aim of clustering, in many applications (e.g., market segmentation) very small clusters are not very useful, and cluster sizes should optimally be close to uniform. This is measured by the well known ``Entropy'' (\cite{shannon1948mathematical}): 
	\begin{displaymath}
		I_{entropy}(\mathcal{C}) = - \sum_{k=1}^{K} \frac{n_{k}}{n} \log(\frac{n_{k}}{n}).
	\end{displaymath}
Large values are good. 
\end{description}
\cite{hennig2017cluster} proposed some more indexes, particularly for measuring within-cluster density decline from the density mean, similarity to a within-cluster uniform or Gaussian distributional shape, and quality of the representation of clusters by their centroids.

\subsection{Stability}
\label{subsec:stability}
Clusterings are often interpreted as meaningful if they can be generalised as stable substantive patterns. Stability means that they can be replicated on different data sets of the same kind. Without requiring that new independent data are available, this can be assessed by resampling methods such as cross-validation and bootstrap. We review two approaches that have been proposed in the literature to measure stability. There they were proposed for estimating the number of clusters on their own, but this is problematic. Whereas it makes sense to require a good clustering to be stable, it cannot be ruled out that an undesirable clustering is also stable. For example, in a data set with four clearly separated clusters, two well separated pairs of clusters may give rise to a potentially even more stable two-cluster solution. We therefore consider these indexes as measuring an isolated aspect of cluster quality to be used in composite indexes. Stability is often of interest on top of whatever criterion characterises the cluster shapes of interest. For example, in applications that require high within-cluster homogeneity, adding a stability criterion can avoid that the data set is split up into too small clusters.  

\begin{description}

	\item[\textbf{Prediction strength (PS):}]
The prediction strength was proposed by \citet{tibshirani2005cluster} for estimating the number of clusters. The data set is split into halves (\cite{tibshirani2005cluster} consider splits into more than two parts as well but settle with halves eventually), say $X_{[1]}$ and $X_{[2]}$. Two clusterings are obtained on these two parts separately with the selected clustering technique and a fixed number of clusters $K$. Then the points of $X_{[2]}$ are classified to the clusters of $X_{[1]}$ in some way. The same is done with the points of $X_{[1]}$ relative to the clustering on $X_{[2]}$. For any pair of observations in the same cluster in the same part, it is then checked whether or not they are predicted to be in the same cluster by the clustering on the other half. This can be repeated for various ($A$) splits of the data set. The prediction strength is then defined as the average proportion of correctly predicted co-memberships for the cluster that minimises this proportion. Formally, 
\begin{align*}
	I_{PS}(\mathcal{C}) &= \frac{1}{2A} \sum_{a=1}^A \sum_{t=1}^{2} \left\{ \min_{1 \leq k \leq K} \left( \frac{m_{kat}}{n_{kat}(n_{kat}-1)} \right) \right\},\\
        m_{kat} &= \sum_{x_i\neq x_{i^{'}} \in C_{kat}} \mathbf{1} \left( l_{i'at}= l_{iat}^{*} \right),\ 1\le k\le K,
\end{align*}
where $C_{kat}$ is cluster $k$ computed on the data half $X_{[t]}$ in the $a$th split, $n_{kat}=|C_{kat}|$ is its number of observations, $L_{at} = \left( l_{g(1)at}, \ldots, l_{g(n/2)at} \right)$ are the cluster indicators of the clustering of $X_{[t]}$ in the $a$th split. $g(1),\ldots,g(n/2)$ denote the indexes of observations belonging to that half, 
assuming for ease of notation that $n$ is even. $L_{at}^{*} = \left\{ l_{g(1)at}^{*}, \ldots, l_{g(n/2)at}^{*} \right\}$ are the clusters of the clustering of the other half $X_{[2-t]}$ in the $s$th split, to which the observations of $X_{(t)}$ are classified.

Unlike the indexes listed in Sections \ref{subsec:popular_cquality} and \ref{subsec:aspect_cquality}, $I_{PS}$ depends on the clustering method applied to arrive at the clusterings $\mathcal{C}$, because stability is evaluated comparing clusterings computed using the same method. Furthermore, $I_{PS}$ requires a supervised classification method to classify the observations in one half of the data set to the clusters computed on the other half. \cite{tibshirani2005cluster} propose classification of observations in one half to the closest cluster centroid in the clustered other half of the data set. This is the same classification rule that is implicitly used by $K$-means and PAM clustering, and therefore it is suitable for use together with these clustering methods. But it is inappropriate for some other clustering methods such as Single Linkage or Gaussian mixture model-based clustering with flexible covariance matrices, in which observations can be assigned to clusters with far away centroids in case of either existence of linking points (Single Linkage) or a within-cluster covariance matrix with large variation in the direction between the cluster centroid and the point to be classified (Gaussian model-based clustering). The classification method used for the prediction strength should be chosen based on the cluster concept formalised by the clustering method in use. Table~\ref{tab:classfnp} lists some classification methods that are associated with certain clustering methods, and we use them accordingly.

\begin{table*}[tb]
	\centering
	\renewcommand{\arraystretch}{1.75}
	\caption{Methods for supervised classification associated to clustering methods. Notation: $a$ refers to the data split, $t$ refers to the data set half (observations of $X_{[t]}$ are classified to clusters of $X_{[2-t]}$), $\mathbf{m}_{ka(2-t)}$ is the centroid and $n_{ka(2-t)}$ the number of points of cluster $k$ in the data set $X_{[2-t]}$, which may depend on the clustering method. For $K$-means and Ward it is the cluster mean, for PAM the medoid minimising the sum of distances to the other points in the cluster. For QDA and LDA, $\delta_{ka(2-t)}(x)=\pi_{ka(2-t)} \left(-\frac{1}{2}\log(|\mathbf{\Sigma}_{ka(2-t)}|)-(x-\mathbf{\mu}_{ka(2-t)})'\mathbf{\Sigma}_{ka(2-t)}^{-1}(x-\mathbf{\mu}_{ka(2-t)})\right)$, where $\pi_{ka(2-t)}$ is the relative frequency, $\mathbf{\Sigma}_{ka(2-t)}$ is the sample covariance matrix (for LDA pooled over all $k$) and $\mathbf{\mu}_{ka(2-t)}$ is the sample mean of cluster $k$.}
	\begin{tabular}{l|c|c}
		Classification method & $l_{iat}^{*}$ & Clustering method\\		
		\hline
		Nearest centroid & $\argmin\limits_{1 \leq k \leq K} d(x_i,\mathbf{m}_{ka(2-t)})$ & $K$-means, PAM, Ward\\
		
		Nearest neighbour & $\argmin\limits_{1 \leq k \leq K} \left( \min_{l_{ja(2-t)}=k} d(x_{i}, x_{j})  \right) $ & Single linkage\\
		
		Furthest neighbour & $\argmin\limits_{1 \leq k \leq K} \left( \max_{l_{ja(2-t)}=k} d(x_{i}, x_{j}) \right) $ & Complete linkage\\
		
		Average dissimilarity & $\argmin\limits_{1 \leq k \leq K} \left( \frac{1}{n_{ka(2-t)}} \sum_{l_{ja(2-t)}=k} d(x_{i}, x_{j})  \right) $ & Average Linkage\\
		
		QDA (or LDA) & $\argmin\limits_{1 \leq k \leq K} \left\{ \delta_{ka(2-t)}(x_{i}) \right\}$ & Gaussian model-based\\
	\end{tabular}
	\label{tab:classfnp}
\end{table*}

Realising that high values of the prediction strength are easier to achieve for smaller numbers of clusters, \cite{tibshirani2005cluster} recommend as estimator for the number of clusters the largest $k$ so that the prediction strength is above 0.8 or 0.9. For using the prediction strength as one of the contributors to a composite index, such a cutoff is not needed.

\item[\textbf{A bootstrap method for measuring stability}] \textbf{(Bootstab,} \citet{fang2012selection}): Similar to the prediction strength, also here the data are resampled, clusterings are generated on the resampled data by a given clustering method with fixed number of clusters $K$. The points in the data set that were not resampled are classified to the clusters computed on the resampled data set by a supervised classification method as listed in Table \ref{tab:classfnp}, and for various resampled data sets the resulting classifications are compared.
  
Here, $A$ times two bootstrap samples are drawn from the data with replacement. Let $X_{[1]},\ X_{[2]}$ the two bootstrap samples in the $a$th bootstrap iteration. For $t=1, 2,$ let $L_{a}^{(t)} = \left( l_{1a}^{(t)}, \ldots, l_{na}^{(t)} \right)$ based on the clustering of $X_{[t]}$. This means that for points $x_i$ that are resampled as member of $X_{[t]}$, $l_{ia}^{(t)}$ is just the cluster membership indicator, whereas for points $x_i$ not resampled as member of $X_{[t]}$, $l_{ia}^{(t)}$ indicates the cluster on $X_{[t]}$ to which $x_i$ is classified using a suitable method from Table \ref{tab:classfnp}.
The Bootstab index is
\begin{displaymath}
	I_{Boot}(\mathcal{C}) = \frac{1}{A} \sum_{a=1}^{A} \left\{ \frac{1}{n^2} \sum_{i,i'} \left|f_{ii^{'}a}^{(1)} - f_{ii^{'}a}^{(2)}\right| \right\},
\end{displaymath}
where for $t=1,2$,
\begin{displaymath}
	f_{ii^{'}a}^{(t)} = \mathbf{1} \left( l_{i'a}^{(t)}= l_{ia}^{(t)} \right),
\end{displaymath}
indicating whether $x_i$ and $x_{i'}$ are in or classified to the same cluster based on the clustering of $X_{[t]}$. $I_{Boot}$ is a percentage of pairs that have different ``co-membership'' status based on clusterings on two bootstrap samples. Small values of $I_{Boot}$ are better. \cite{fang2012selection} suggest to choose the number of clusters by minimising  $I_{Boot}$. Without proof they imply that this method is not systematically biased in favour of smaller numbers of clusters.

\end{description}

\section{Aggregation and Calibration for Definition of a Composite Index}
\label{sec:aggregation_indexes}

As discussed earlier, different aspects of cluster quality are typically relevant in different applications. From a list of desirable characteristics of clusters in a given application a composite index can be constructed as a weighted mean of indexes that measure the specific characteristics of interest. This index can then be optimised. We will here assume that for all involved indexes larger values are better (involved indexes for which this is not the case can be multiplied by $-1$ to achieve this). For selected indexes $I_{1}, \ldots, I_{s}$ with weights $w_{1}, \ldots, w_{s} > 0$:
\begin{equation}
	\mathcal{A}(\mathcal{C}) = \frac{\sum_{j=1}^{s} w_{j} I_{j}(\mathcal{C})}{\sum_{j=1}^{s} w_{j}}.
	\label{eq:aggregation_indexes}
\end{equation} 
In order to choose the weights in a given application, it would be useful if it were possible to interpret the weights in terms of the relative importance of the desirable characteristics. This requires that the values of the different $I_{1}, \ldots, I_{s}$ can be meaningfully compared; a loss of 0.3, say, in one index should in terms of overall quality be offset by an improvement of 0.3 in another index of the same weight. For the indexes defined in Section \ref{sec:internal_validation_indexes}, this does not hold. Value ranges and variation will potentially differ strongly between indexes.  

Here we transform the indexes relative to their expected variation over clusterings of the same data. This requires a random routine to generate many clusterings on the data set. Note the difference to standard thinking about random variation where the data is random and a method's results are fixed, whereas here the data are treated as fixed and the clusterings as random. For transforming the indexes relative to these, standard approaches such as $Z$-scores or range transformation can be used. 

The random clusterings should make some sense; one could just assign points to clusters in a random fashion, but then chances are that most index values from a proper application of an established clustering method will be clearly better then those generated from the random clusterings, in which case transforming the indexes relative to the random clusterings is not appropriate. On the other hand the algorithms to generate random clusterings need to provide enough variation for their distribution to be informative. Furthermore, for the aim of making the indexes comparable, random clusterings should optimally not rely on any specific cluster concept, given that different possible concepts are implied by the different indexes. 

In order to generate random clusterings that are sensible, though, a certain cluster concept or definition is required. We treat this problem by proposing four different algorithms for generating random clusterings that correspond to different cluster concepts, more precisely to $K$-centroids (clusters for which all points are close to the cluster centroid), single linkage (connected clusters of arbitrary shape), complete linkage (limiting the maximum within-cluster dissimilarity), and 
average linkage (a compromise allowing for flexible shapes but not for too many large within-cluster dissimilarities or too weak connection). 

The number of cluster $K$ is always treated as fixed for the generation of random clusterings. The same number of random clusterings should be generated for each $K$ from a range of interest. This also allows to assess whether and to what extent certain indexes are systematically biased in favour of small or large $K$. 

\subsection{Random $K$-centroids}
\label{subsec:random_k_centroid}
Random $K$-centroids works like a single step of Lloyd's classical $K$-means algorithm (\cite{Lloyd82}) with random initialisation. Randomly select $K$ cluster centroids from the data points, and assign every observation to the closest centroid, see Algorithm \ref{alg:random_k_centroid}. 

\SetAlFnt{\small\sffamily}
\IncMargin{0.1em}
\begin{algorithm}[ht]
	\DontPrintSemicolon
	\SetKwFunction{Output}{Output}
	\SetKwInOut{Input}{input}\SetKwInOut{Output}{output}
	\BlankLine
	\Input{ $\mathcal{X} = \left\{ x_{1}, \ldots, x_{n} \right\}$ (objects), $\mathbf{D} = \left(d(x_{i}, x_{j})\right)_{i,j=1,\ldots, n}$ (dissimilarities),  
	$K$ (number of clusters)}
	\Output{$\mathcal{L}=\left(l_{1}, l_{2}, \ldots, l_{n} \right)$ (cluster labels)}
	\BlankLine
	\textbf{INITIALISATION: } \;
Choose $K$ random centroids 
${\cal S}\leftarrow \{s_{1}, s_{2}, \ldots, s_{K}\}$
according to the uniform distribution over subsets of size $K$ from $\mathcal{X}$ \;


	\For{$i \leftarrow 1$ \KwTo $n$}{
	\# Assign every observations to the closest centroid: \;
	$l_{i} = \argmin\limits_{1 \leq k \leq K} d(x_{i}, s_{k}) $, $i \in N_{n}$
	}
	\Return{$\mathcal{L}$, indexing clustering $\mathcal{C}_{rK-cen}(\mathcal{S})$}

	\BlankLine
	\caption{Random $K$-centroids algorithm}
	\label{alg:random_k_centroid}
\end{algorithm}

\subsection{Random $K$-linkage methods}
\label{subsec:random_k_neighbour}

The three algorithms random $K$-single, random $K$-complete, and random $K$-average are connected to the basic hierarchical clustering methods single, complete, and average linkage. As for random $K$-centroids, the clustering starts from drawing $K$ initial observations at random, forming $K$ one-point clusters. Then clusters are grown by adding one observations at a time to the closest cluster, where closeness is measured using the dissimilarity to the closest neighbour (single), to the furthest neighbour (complete), or the average of dissimilarities (average), see Algorithm~\ref{alg:random_k_neighbour}.

\SetAlFnt{\small\sffamily}
\IncMargin{0.25em}
\begin{algorithm}[!htp]
	\DontPrintSemicolon
	\SetKwInOut{Input}{input}
	\SetKwInOut{Output}{output}
	\BlankLine
	\Input{	$\mathcal{X} = \left\{ x_{1}, \ldots, x_{n} \right\}$ (objects), $\mathbf{D} = \left(d(x_{i}, x_{j})\right)_{i,j=1,\ldots, n}$ (dissimilarities),  
	$K$ (number of clusters)}
	\Output{$\mathcal{C} = \left\{ C_{1}, \ldots, C_{K} \right\}$ (set of clusters)}
	
	\BlankLine
	\textbf{INITIALISATION: } \;
Choose $K$ random initial points
$\mathcal{S} \leftarrow \{s_{1}, s_{2}, \ldots, s_{K}\}$
according to the uniform distribution over subsets of size $K$ from $\mathcal{X}$ \;	
Initialise clusters $\mathcal{C}(\mathcal{S}) = \left\{ C_{1}, \ldots, C_{K} \right\} \leftarrow \left\{ \left\{ s_{1} \right\}, \ldots, \left\{ s_{K} \right\}  \right\}$ \;
	$t \leftarrow 1$; $\mathcal{R}= \mathcal{X} \setminus \mathcal{S}$; \;
        $\mathbf{D}^{(t)}=\left(d^{(t)}(x,C)\right)_{x\in\mathcal{R}, C\in\mathcal{C}},\ d^{(t)}(x,C_j)=d(x,s_j),\ j=1,\ldots,K$ \;		
	
	\Repeat{$\mathcal{R} = \emptyset$}{
	\textbf{STEP 1:} \;
	\hspace{10mm} 
$(g,h) \leftarrow \argmin\limits_{x_{g} \in \mathcal{R}, C_{h} \in \mathcal{C}} d^{(t)}(x_{g}, C_{h})$ \;
	
	\textbf{STEP 2:} \;
\hspace{10mm} $C_{h} \leftarrow C_{h} \cup \{x_g\}$ with $\mathcal{C}$ updated accordingly, \;
\hspace{10mm} $ \mathcal{R} \leftarrow \mathcal{R} \setminus \left\{ x_{g} \right\}$ \;
	
	\textbf{STEP 3:} \;
	\ForEach{$x \in \mathcal{R},\ C_j\in \mathcal{C}$}
	{ Update $d^{(t+1)}(x, C_j) \leftarrow d^{(t)}(x, C_j)$ for $j\neq h;$ and\;
	\hspace{1mm} \textbf{Random $K$-single:} $d^{(t+1)}(x, C_h) \leftarrow\min_{y\in C_h} d^{(t)}(x, y)$, \;
	
	\hspace{1mm} \textbf{Random $K$-complete:} $d^{(t+1)}(x, C_h) \leftarrow\max_{y\in C_h} d^{(t)}(x, y)$, \;	
	\hspace{1mm} \textbf{Random $K$-average:} $d^{(t+1)}(x, C_h) \leftarrow\frac{1}{|C_h|}\sum_{y\in C_h}d^{(t)}(x, y)$.

	}
	$t \leftarrow t+1$ \;
	}
	\Return{$\mathcal{C}$, denoted $\mathcal{C}_{rKsin}(\mathcal{S}),\ \mathcal{C}_{rKcom}(\mathcal{S}),$ or $\mathcal{C}_{rKave}(\mathcal{S})$}	
		
	\BlankLine
	\caption{Random $K$-single / complete / average linkage algorithms}
	\label{alg:random_k_neighbour}
\end{algorithm}

\subsection{Calibration}
\label{subsec:calibration}

The random clusterings can be used in different ways to calibrate the clustering validity indexes. For given $B$ and any value of the number of clusters $K\in \left\{ 2, \ldots, K_{max} \right\}$ of interest, $4B + R_K$ clusterings and corresponding index values are computed, where $R_K$ is the number of ``genuine'' clusterings for given $K$, $\mathcal{C}_1,\ldots,\mathcal{C}_{R_K},$ to be validated originally, i.e., those generated by ``proper'' clustering methods (as opposed to the random clusterings generated for calibration), and $\mathcal{S}_1,\ldots,\mathcal{S}_{4B}$ are initialisation sets of size $K$: 
\[ \begin{split}
\mathcal{C}_{Kcol} = &\left(\mathcal{C}_{K:1}, \ldots, \mathcal{C}_{K:4B+R}\right) = \\
& \left(\mathcal{C}_{rKcen}(\mathcal{S}_{1}), \ldots, \mathcal{C}_{rKcen}(\mathcal{S}_{B}),\right. \\
& \mathcal{C}_{rKsin}(\mathcal{S}_{B+1}), \ldots, \mathcal{C}_{rKsin}(\mathcal{S}_{2B}), \\
& \mathcal{C}_{rKcom}(\mathcal{S}_{2B+1}), \ldots, \mathcal{C}_{rKcom}(\mathcal{S}_{3B}), \\
& \mathcal{C}_{rKave}(\mathcal{S}_{3B+1}), \ldots, \mathcal{C}_{rKave}(\mathcal{S}_{4B})), \\
& \left.\mathcal{C}_1,\ldots,\mathcal{C}_{R_K}\right),
			\end{split}	\]
			
with further notation as in Algorithms~\ref{alg:random_k_centroid} and \ref{alg:random_k_neighbour}. $\mathcal{C}_{Kcol}$ stands for the collection of clustering validity indexes computed from the real and random clustering algorithms. 

There are two possible approaches to calibration:

\begin{itemize}
	\item Indexes for clusterings with $K$ clusters can be calibrated relative to proper and random clusterings for the same $K$ only. Indexes are assessed relative to what is expected for the same $K$, with a potential to correct systematic biases of indexes against small or large $K$.
	\item Indexes for all clusterings can be calibrated relative to genuine and random clusterings for all values of $K$ together. Here, raw index values are compared over different values for $K$. This cannot correct systematic biases of indexes, but may be suitable if the raw index values appropriately formalise what is required in the application of interest, and that indexes that systematically improve for larger $K$ (such as average within-cluster distances) are balanced by indexes that favour a smaller number (such as separation or prediction strength).  	
\end{itemize}
For the second approach, $\mathcal{C}_{col}=\bigcup_{K=2}^{K_{max}}\mathcal{C}_{Kcol}$, which is used below instead of $\mathcal{C}_{Kcol}$.

There are various possible ways to use the collection of random clusterings for standardisation, for a given index $I(\mathcal{C})$ for a given clustering $\mathcal{C}$ with $|\mathcal{C}|=K$. We use $Z$-score standardisation here:
	\begin{align*}
	\scriptstyle
		I^{Z-score}(\mathcal{C}) = \frac{I(\mathcal{C}) - \mathbf{m}(\mathcal{C}_{Kcol})}{ \sqrt{\frac{1}{|\mathcal{C}_{Kcol}|-1} \sum_{\mathcal{C}^*\in \mathcal{C}_{Kcol}} \left( I(\mathcal{C}^*) - \mathbf{m}(\mathcal{C}_{Kcol}) \right)^{2}}},
	\end{align*}
where 
\begin{align*}
	\scriptstyle
	\mathbf{m}(\mathcal{C}_{Kcol}) = \frac{1}{|\mathcal{C}_{Kcol}|} 	\sum_{\mathcal{C}^*\in \mathcal{C}_{Kcol}} I(\mathcal{C}^*).
\end{align*}

Further options are for example standardisation to range $(0,1)$, or transformation of all values in the collection to ranks, which would lose the information in the precise values, but is less affected by outliers.

%

\begin{figure*}[!htbp]
\begin{center}
\includegraphics[width=0.485\textwidth]{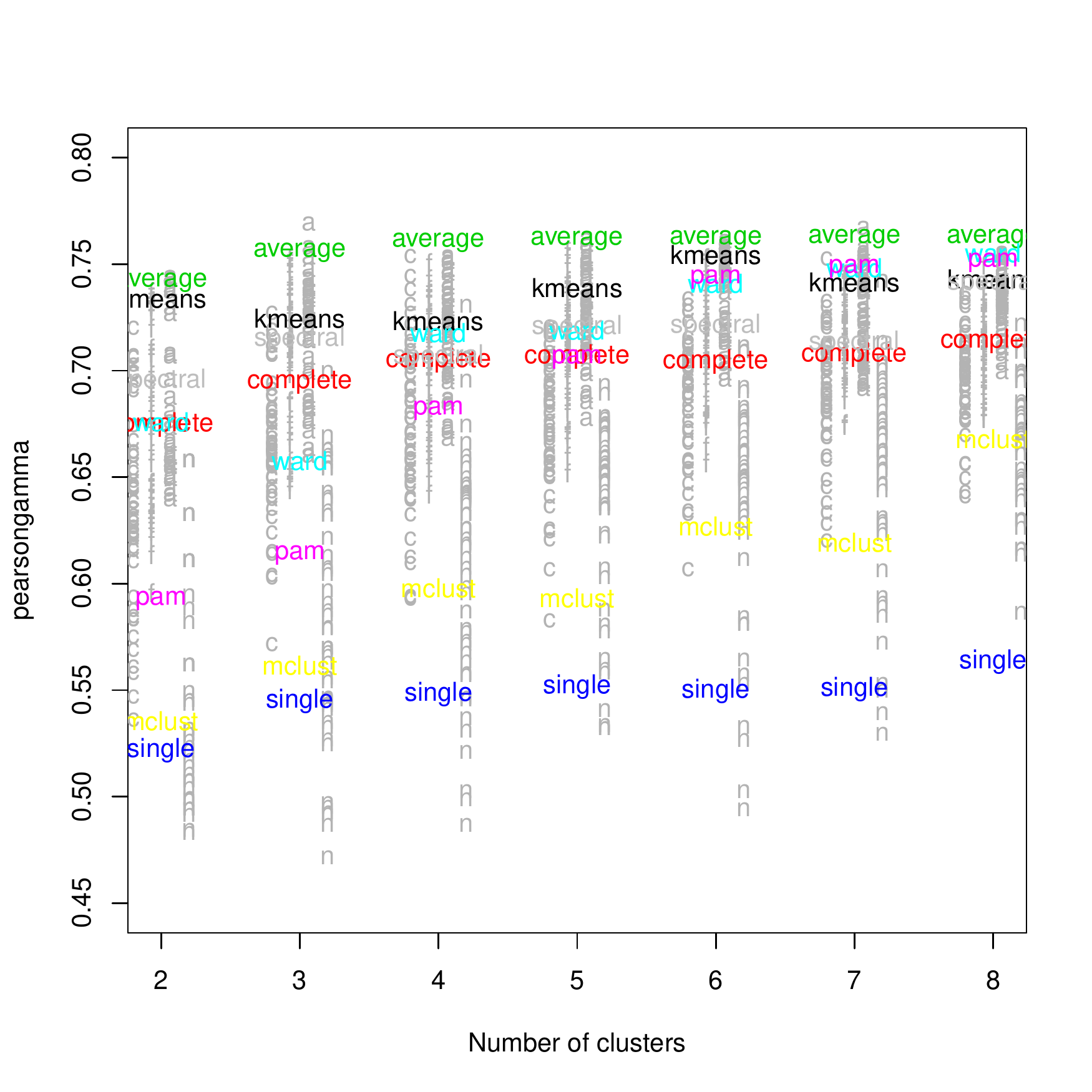}
\includegraphics[width=0.485\textwidth]{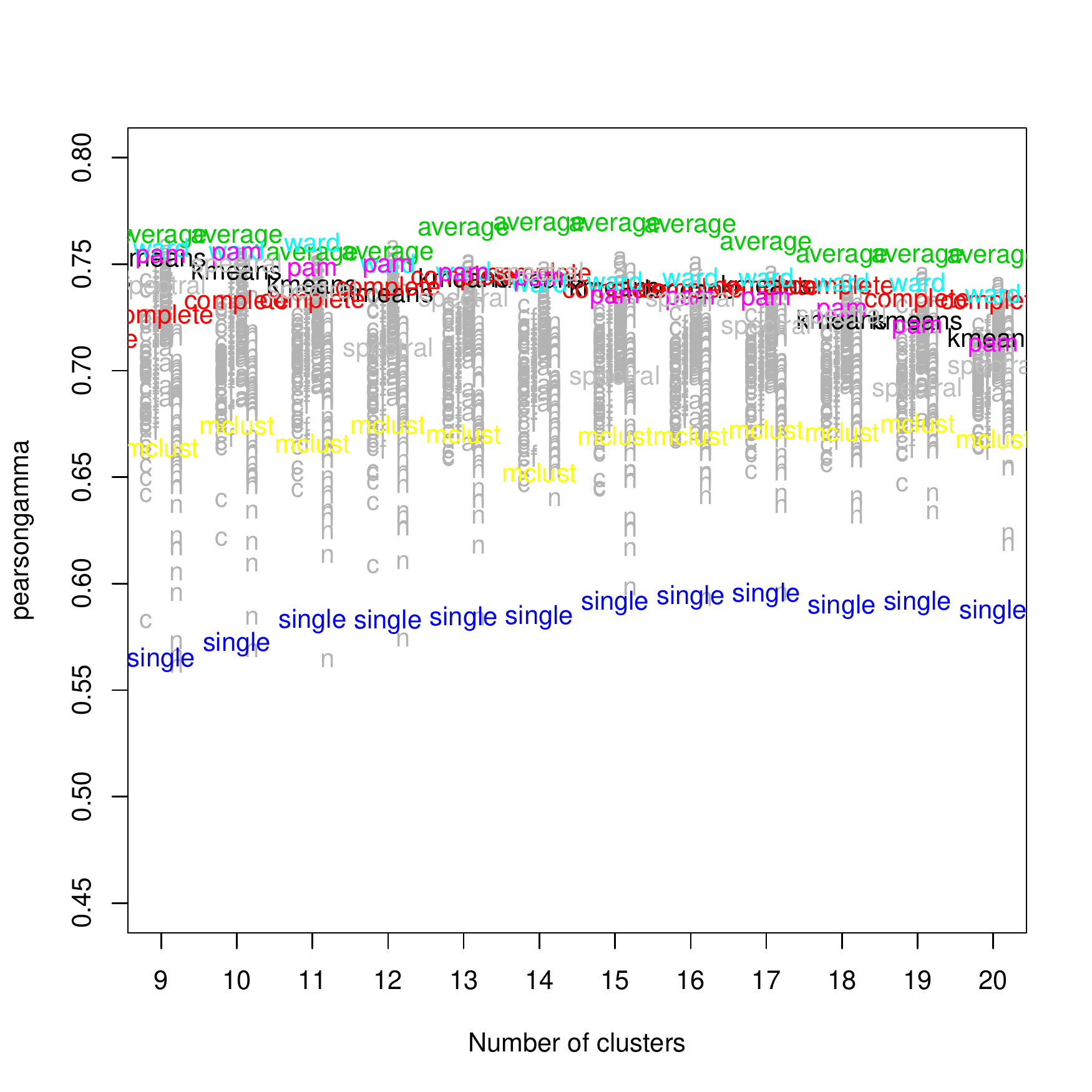}
\end{center}
\caption{Values of the Pearson$\Gamma$ index for eight clustering methods on Movement data for numbers of clusters 2-8 (left side) and 9-20 (right side) together with values achieved by random clusterings, denoted by grey letters ``c'' 
(random $K$-centroids), ``f'' (random furthest neighbour/complete), ``a'' (random average neighbour), ``n'' (random nearest neighbour/single), from left to right for each number of clusters).}
\label{fmovementcalibration}
\end{figure*}

Figure \ref{fmovementcalibration} serves to illustrate the process. It shows the uncalibrated values of Pearson$\Gamma$ achieved by the various clustering methods on the Movement data set (see Section \ref{smovement}). For aggregation with other indexes, the Pearson$\Gamma$ index is calibrated by the mean and standard deviation taken from all those clusterings, where the random clusterings indicate the amount of variation of the index on these data. It can also be seen that the different methods to generate random clusterings occasionally result in quite different distributions, meaning that taken all together they give a more comprehensive expression of the variation less affected by specific cluster concepts. 

One may wonder whether random methods are generally too prone to deliver nonsense clusterings that cannot compete or even compare with the clusterings from the ``proper'' clustering methods, but it can be seen in Figure \ref{fmovementcalibration}, and also later in Figures \ref{fwineresults1}, \ref{fwineresults2}, \ref{fseedsresults}, \ref{fmovementresults}, and \ref{fbundestagresults}, that the random clusterings often achieve better index values than at least the weakest ``proper'' clusterings if not even the majority of them; note that only in some of these figures random clustering results are actually shown, but more can be inferred from some $Z$-score transformed values not being significantly larger than 0. This also happens for validation indexes already in the literature.

\section{Applications and experiments}
\label{sec:numerical_experiment}
\subsection{General approach and basic composite indexes}
The approach presented here can be used in practice to design a composite cluster validity index based on indexes that formalise characteristics that are required in a specific application using (\ref{eq:aggregation_indexes}). This can be used to compare clusterings generated by different clustering methods with different numbers of clusters (or other required parameter choices) and to pick an optimal solution if required; it may be of interest to inspect not only the best solution.

By choosing weights for the different indexes different situations can be handled very flexibly using background and subject matter information as far as this is available. However, in order to compare the approach with existing indexes and to investigate its performance in more generality, such flexibility cannot be applied. In particular, it is not possible to choose index weights from background information for simulated data. For this reason we proceed here in a different way. 

\cite{hennig2017cluster} presented an example in which indexes are chosen according to subject matter knowledge about the characteristics of the required clustering. This was about biological species delimitation for a data set with genetic information on bees. The chosen indexes were $I_{ave.wit}$ (individuals within species should be genetically similar), $I_{widest.gap}$ (a genetic gap within a species runs counter to general genetic exchange within a species), $I_{sep.index}$ (species should be genetically separated from other species), and $I_{Pearson\Gamma}$ (generally species should represent the genetic distance structure well). Other aspects such as entropy or representation of individuals by centroids were deemed irrelevant for the species delimitation task.

Here we investigate a more unified application of the methodology to data sets with known true clusters, either simulated or real. This runs counter to some extent to the philosophy of \cite{hennig15true}, where it is stated that even for data sets with given true clusters it cannot be taken for granted that the ``true'' ones are the only ones that make sense or could be of scientific interest. However, it is instructive to see what happens in such situations, and it is obviously evidence for the usefulness of the approach if it is possible to achieve good results recovering such known true clusters. 

We simulated scenarios and applied methods to some real data sets using several combinations of indexes, in order to find combinations that have a good overall performance. One should not expect that a single composite index works well for all data sets, because ``true'' clusters can have very different characteristics in different situations. We found, however, that there are two composite indexes that could be used as some kind of basic toolbox, namely ${\cal A}_1$, made up of $I_{ave.wit}, I_{Pearson\Gamma}$ and $I_{Boot}$, and ${\cal A}_2$, made up of $I_{sep.index}, I_{widest.gap},$ and $I_{Boot}$ (all with equal weights). Calibration was done as explained in Section \ref{subsec:calibration} with $B=100$.

${\cal A}_1$ emphasises cluster homogeneity by use of $I_{ave.wit}$. $I_{Pearson\Gamma}$ supports small within-cluster distances as well but will also prefer distances between clusters to be large, adding some protection against splitting already homogeneous clusters, which could happen easily if $I_{ave.wit}$ would be used without a corrective. Stability as measured by $I_{Boot}$  is of interest in most clustering tasks, and is another corrective against producing too small spurious clusters. 

${\cal A}_2$ emphasises cluster separation by use of $I_{sep.index}$. Whereas $I_{sep.index}$ looks at what goes on between clusters, $I_{within.gap}$ makes sure that gaps within clusters are avoided; otherwise one could achieve strong separation by only isolating the most separated clusters and leaving clusters together that should intuitively better be split. Once more $I_{Boot}$ is added for stability. At least one of  ${\cal A}_1$ and ${\cal A}_2$ worked well in all situations, although neither of these (and none of the indexes already in the literature) works well universally, as expected.

In our experiments we found that overall  $I_{Boot}$ worked slightly better than $I_{PS}$ for incorporating stability in an composite index, aggregating over all numbers of clusters $K$ together worked better at least for ${\cal A}_1$ and ${\cal A}_2$ than aggregating separately for separate $K$ (this is different for some other composite indexes; also separate aggregation may be useful where aggregating over all $K$ leads to ``degenerate'' solutions at the upper or lower bound of $K$). Z-score standardisation looked overall slightly preferable to other standardisation schemes, but there is much variation and not much between them overall. We only present selected results here, particularly not showing composite indexes other than ${\cal A}_1$ and ${\cal A}_2$, standardisation other than Z-score, and aggregation other than over all $K$ together. Full results are available from the authors upon request. There may be a certain selection bias in our results given that ${\cal A}_1$ and ${\cal A}_2$ were selected based on the results of our experiments from a large set of possible composite indexes. Our aim here is not to argue that these indexes are generally superior to what already exists in the literature, but rather to make some well founded recommendations for practice and to demonstrate their performance characteristics. ${\cal A}_1$ and ${\cal A}_2$ were not selected by formal optimisation over experimental results, but rather for having good interpretability of the composite indexes (so that in a practical situation a user can make a decision without much effort), and the basic tension in clustering between within-cluster homogeneity and between-clusters separation.

Note that composite indexes involving entropy could have performed even better for the simulated and benchmark data sets with given true clusters below, because the entropy of the given true clusters is in most cases perfect or at least very high. But involving Entropy here seemed unfair to us, already knowing the true clusters' entropy for the simulated and benchmark data sets, whereas in reality a high entropy cannot be taken for granted. Where roughly similar cluster sizes are indeed desirable in reality, we recommend to involve entropy in the composite indexes.    

Results for cluster validation and comparison of different numbers of clusters generally depend on the clustering algorithm that is used for a fixed number of clusters. Here we applied 8 clustering algorithms (Partitioning Around Medoids (PAM), $K$-means, Single Linkage, Complete Linkage, Average Linkage, Ward's method, Gaussian Model based clustering - mclust, 
Spectral Clustering; for all of these standard R-functions with default settings were used). All were combined with the validity indexes CH, ASW, Dunn, Pearson$\Gamma$, CVNN (with $\kappa=10$), and the stability statistics PS and Bootstab with $A=50$. PS was maximised, which is different from what is proposed in \cite{tibshirani2005cluster}, where the largest number of clusters is selected for which PS is larger than some cutoff value. For the recommended choices of the cutoff, in our simulations many data sets would have produced an estimator of 1 for the number of clusters due to the lack of any solution with $K\ge 2$ and large enough PS, and overall results would not have been better. One popular method that is not included is the BIC for mixture models (\cite{fraley2002}). This may have performed well together with mclust in a number of scenarios, but is tied to mixture models and does not provide a more general validity assessment.

\subsection{Simulation study}
\begin{figure*}[!htbp]
	\begin{center}
  	\subfloat[\tiny PAM, $K=3$ ($ARI = 0.990$)]{\includegraphics[width=5.5cm, height=5cm]{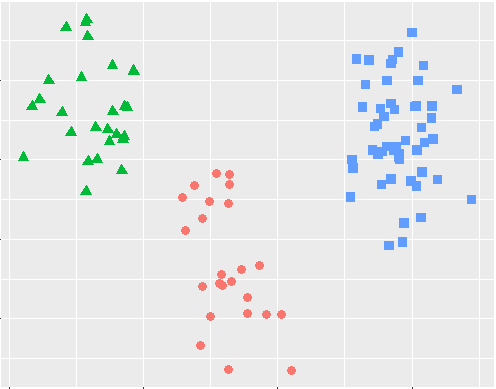}}
 	\hfill
  	\subfloat[\tiny mclust, $K=4$ ($ARI = 0.955$)]{\includegraphics[width=5.5cm, height=5cm]{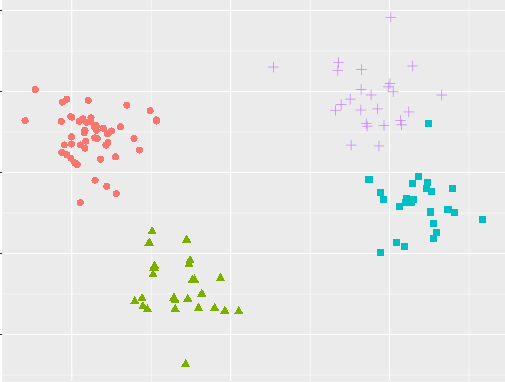}}
  	\hfill
  	\subfloat[\tiny mclust, $K=4$ ($ARI = 0.834$)]{\includegraphics[width=5.5cm, height=5cm]{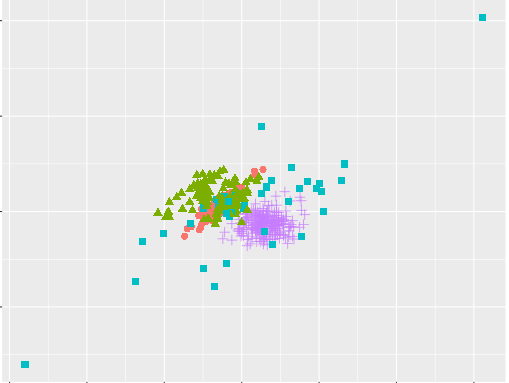}}
  	\hfill
  	\subfloat[\tiny Complete linkage, $K=2$ ($ARI = 1.000$)]{\includegraphics[width=5.5cm, height=5cm]{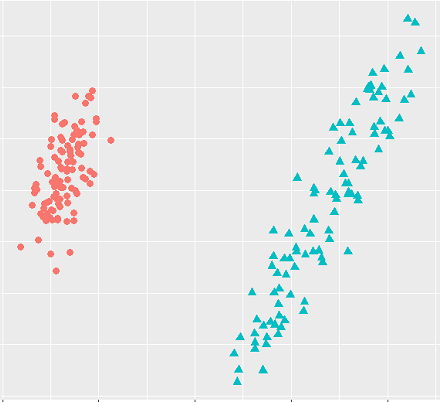}}
  	\hfill
  	\subfloat[\tiny Single linkage, $K=2$ ($ARI = 1.000$)]{\includegraphics[width=5.5cm, height=5cm]{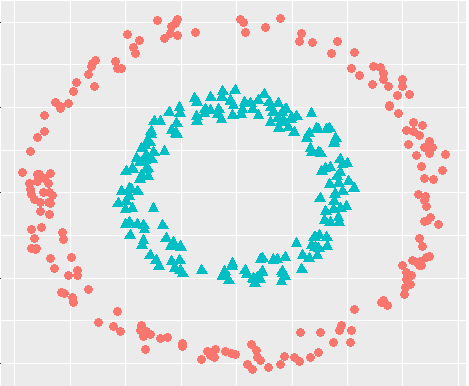}}
	\hfill
	\subfloat[\tiny Spectral clustering, $K=2$ ($ARI = 1.000$)]{\includegraphics[width=5.5cm, height=5cm]{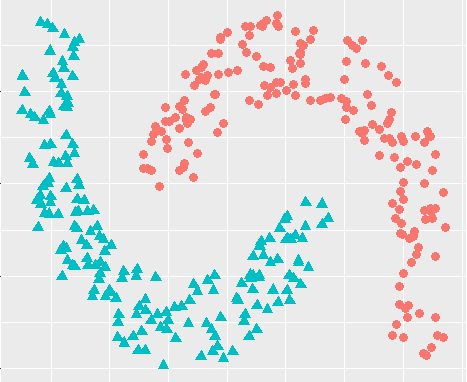}}
	\end{center}
  	\caption{Data sets produced by Scenarios 1-6 (for the more than 2-dimensional data sets from Scenarios 2-4 principal components are shown) with the respective clusterings that achieve highest ARI on the true number of clusters.}

	\label{fig:simulated_data_pca2}
\end{figure*}

For comparing the composite indexes ${\cal A}_1$ and ${\cal A}_2$ with the other validity and stability indexes, data were generated from six different scenarios, covering a variety of clustering problems (obviously we cannot claim to be exhaustive). 50 data sets were generated from each scenario. Scenario 1, 2, and 4 are from \citet{tibshirani2005cluster}, scenario 3 from \citet{hennig2007}, and scenarios 5 and 6 from the R-package \texttt{clusterSim}, \citet{walesiak2011clustersim}. Figure~\ref{fig:simulated_data_pca2} shows data from the six scenarios. 


\begin{itemize}
 	\item \textbf{Scenario 1 (Three clusters in two dimensions):} Clusters are normally distributed with 25, 25, and 50 observations, centred at $(0,0)$, $(0,5)$, and $(5, -3)$ with identity covariance matrices.  
 	
 	\item \textbf{Scenario 2 (Four clusters in 10 dimensions):} Each of four clusters was randomly chosen to have $25$ or $50$ normally distributed observations, with centres randomly chosen from $N(0, 1.9I_{10})$. Any simulation with minimum between-cluster distance less than $1$ was discarded in order to produce clusters that can realistically be separated. 
 	
 	\item \textbf{Scenario 3 (Four or six clusters in six dimensions with mixed distribution types):} 
\citet{hennig2007} motivates this as involving some realistic issues such as different distributional shapes of the clusters and multiple outliers. The scenario has four ``clusters'' and two data subgroups of outliers. There is an ambiguity in cluster analysis regarding whether groups of outliers should be treated as clusters, and therefore the data could be seen as having six clusters as well. Furthermore, there are two ``noise'' variables not containing any clustering information (one ${\cal N}(0,1)$, the other $t_2$), and the clustering structure is defined on the first four dimensions. 

Cluster 1 (150 points): Gaussian distribution with mean vector (0, 2, 0, 2) and covariance matrix 0.1$I_4$. Cluster 2 (250 points): Gaussian distribution with mean vector (3, 3, 3, 3) and a covariance matrix with diagonal elements 0.5 and covariances 0.25 in all off-diagonals. Cluster 3 (70 points): A skew cluster with all four dimensions distributed independently exponentially (1) shifted so that the mean vector is (-1,−1,−1,−1). Cluster 4 (70 points): 4-variate $t_2$-distribution with mean vector (2, 0, 2, 0) and Gaussian covariance matrix 0.1$I_4$ (this is the covariance matrix of the Gaussian distribution involved in the definition of the multivariate t-distribution). Outlier cluster 1  (10 points): Uniform$[−2, 5]$. Outlier cluster 2 (10 points): 4-variate $t_2$-distribution with mean vector (1.5, 1.5, 1.5, 1.5) and covariance matrix (see above) 2$I_4$.

 	\item \textbf{Scenario 4 (Two elongated clusters in three dimensions):} Cluster 1 was generated by setting, for all points, $x_{1} = x_{2} = x_{3} = t$ with $t$ taking on 100 equally spaced values from $-.5$ to $.5$. Then Gaussian noise with standard deviation $.1$ is added to every variable.
Cluster 2 is generated in the same way, except that the value $1$ is then added to each variable.

	\item \textbf{Scenario 5 (Two ring-shaped clusters in two dimensions):} Generated by function \texttt{shapes.circles2} of the R-package \texttt{clusterSim}.  For each point a random radius $r$ is generated (see below), then a random angle $\alpha \sim U[0, 2\pi] $. The point is then $(r\cos(\alpha), r\sin(\alpha))$. Default parameters are used so that each cluster has 180 points. $r$ for the first cluster is from Uniform$[0.75, 0.9]$, for the second cluster from Uniform$[0.35, 0.5]$.
	
	\item \textbf{Scenario 6 (Two moon-shaped clusters in two dimensions):} Generated by function \texttt{shapes.two.moon} of the R-package \texttt{clusterSim}. For each point a random radius $r$ is generated from Uniform$[0.8, 1.2]$, then a random angle $\alpha \sim U[0, 2\pi]$, and the points are $(a+|r\cos(\alpha)|, r\sin(\alpha))$ for the first cluster and $(-|r\cos(\alpha)|, r\sin(\alpha)-b)$ for the second cluster. Default parameters are used so that each cluster has 180 points, $a=-0.4$ and $b=1$.
\end{itemize}

Results are given in Tables~\ref{tab:simulated_data_clust_valid_res} and \ref{tab:simulated_data_clust_valid_res2}. All these results are based on the clustering method that achieved highest Adjusted Rand Index (ARI) for the true number of clusters. This was decided because of the large number of results, and because the clustering method chosen in this way gave the validation methods the best chance to find a good clustering at the true number of clusters. Figure~\ref{fig:simulated_data_pca2} shows these clusterings.

\begin{table*}[!htbp]
	\renewcommand{\arraystretch}{1.25}
	\scriptsize
	\caption{Results of Simulation Study. Numbers are counts out of 50 trials. Counts for estimates larger
than 10 are not displayed. ``*'' indicates true number of clusters.}
	\centering
		\begin{tabular}{l | l | l l l l l l l l l}
		\multirow{2}{*}{\textbf{Validity Index}} & \multirow{2}{*}{\textbf{ARI}} & \multicolumn{9}{c}{\textbf{\underline{Estimate of Number of Clusters}}} \\
		&& $2$ & $3$ & $4$ & $5$ & $6$ & $7$ & $8$ & $9$ & $10$ \\
		\hline 
		\multicolumn{11}{c}{\underline{\textit{Scenario 1 - Three clusters in 2-d - PAM clustering}}} \\  
		CH                  & 0.990 & $0$  & $50*$ & $0$ & $0$ & $0$ & $0$ & $0$ & $0$ & $0$ \\
		ASW                 & 0.961 & $6$  & $44*$ & $0$ & $0$ & $0$ & $0$ & $0$ & $0$ & $0$ \\
		Dunn                & 0.937 & $11$ & $39*$ & $0$ & $0$ & $0$ & $0$ & $0$ & $0$ & $0$ \\
		Pearson $\Gamma$        & 0.990 & $0$  & $50*$ & $0$ & $0$ & $0$ & $0$ & $0$ & $0$ & $0$ \\
		Prediction strength  & 0.966 & $5$  & $45*$ & $0$ & $0$ & $0$ & $0$ & $0$ & $0$ & $0$ \\
		Bootstab      & 0.990 & $0$  & $50*$ & $0$ & $0$ & $0$ & $0$ & $0$ & $0$ & $0$ \\
		CVNN      & 0.990 & $0$  & $50*$ & $0$ & $0$ & $0$ & $0$ & $0$ & $0$ & $0$ \\
		${\cal A}_1$ & 0.990 & $0$ & $50*$ & $0$ & $0$ & $0$ & $0$ & $0$ & $0$ & $0$ \\
		${\cal A}_2$ & 0.942 & $10$ & $40*$ & $0$ & $0$ & $0$ & $0$ & $0$ & $0$ & $0$ \\
		
		\hline 
		\multicolumn{11}{c}{\underline{\textit{Scenario 2 - Four clusters in 10-d - model-based (mclust) clustering}}} \\  
		CH                  & 0.879 & $6$  & $6$  & $38*$ & $0$ & $0$ & $0$ & $0$ & $0$ & $0$ \\
		ASW                 & 0.815 & $ 9$ & $9$  & $32*$ & $0$ & $0$ & $0$ & $0$ & $0$ & $0$ \\
		Dunn                & 0.796 & $12$ & $5$  & $32*$ & $1$ & $0$ & $0$ & $0$ & $0$ & $0$ \\
		Pearson $\Gamma$        & 0.799 & $7$  & $15$ & $28*$ & $0$ & $0$ & $0$ & $0$ & $0$ & $0$ \\
		Prediction strength  & 0.633 & $28$ & $8$  & $14*$ & $0$ & $0$ & $0$ & $0$ & $0$ & $0$ \\
		Bootstab      & 0.749 & $13$  & $5$  & $ 9*$ & $23 $ & $0$ & $0$ & $0$ & $0$ & $0$ \\
		CVNN      & 0.934 & $1$  & $4$  & $45*$ & $0$ & $0$ & $0$ & $0$ & $0$ & $0$ \\
		${\cal A}_1$& 0.930 & $0$ & $4$  & $36*$ & $10$ & $0$ & $0$ & $0$ & $0$ & $0$ \\
		${\cal A}_2$ & 0.709 & $18$ & $11$ & $20*$ & $1$ & $0$ & $0$ & $0$ & $0$ & $0$ \\
		
		\hline 
		\multicolumn{11}{c}{\underline{\textit{Scenario 3 - Four or six clusters in 6-d}}} \\
		\multicolumn{11}{c}{\underline{\textit{with mixed distribution types - model-based (mclust) clustering.}}} \\
		CH                  & 0.567 & $30$  & $11$  & $3*$ & $2$  & $1*$ & $1$ & $0$ & $2$  & $0$  \\
		ASW                 & 0.454 & $35$  & $2$ & $10*$ & $1$  & $2*$ & $0$ & $0$ & $0$  & $0$  \\
		Dunn                & 0.571 & $23$  & $12$  & $4*$  & $5$  & $4*$ & $1$ & $0$ & $0$ & $1$ \\
		Pearson $\Gamma$        & 0.587 & $15$  & $3$ & $18*$ & $8$ & $5*$ & $0$ & $1$ & $0$  & $0$  \\
		Prediction strength  & 0.418 & $39$ & $11$  & $0*$  & $0$  & $0*$ & $0$ & $0$ & $0$  & $0$  \\
		Bootstab      & 0.807 & $2$  & $12$  & $36*$  & $0$  & $0*$ & $0$ & $0$ & $0$  & $0$ \\
		CVNN      & 0.568 & $32$  & $5$  & $3*$ & $2$  & $2*$ & $2$ & $1$ & $1$  & $2$  \\ 
		${\cal A}_1$& 0.788 & $2$ & $1$  & $37*$ & $ 2$ & $4*$ & $0$ & $0$ & $2$ & $2$ \\
		${\cal A}_2$ & 0.739 & $ 1$ & $ 5$ & $26*$ & $4$ & $0*$ & $0$ & $0$ & $3$ & $11$ \\
		\thickhline 
		\end{tabular}
	\label{tab:simulated_data_clust_valid_res}	
\end{table*}

\begin{table*}[!htbp]
	\renewcommand{\arraystretch}{1.25}
	\scriptsize
	\caption{Continuous of Table~\ref{tab:simulated_data_clust_valid_res}}
	\centering
		\begin{tabular}{l | l | l l l l l l l l l}
		\multirow{2}{*}{\textbf{Validity Index}} & \multirow{2}{*}{\textbf{ARI}} & \multicolumn{9}{c}{\textbf{\underline{Estimate of Number of Clusters}}} \\
		&& $2$ & $3$ & $4$ & $5$ & $6$ & $7$ & $8$ & $9$ & $10$ \\
		\hline  
		\multicolumn{11}{c}{\underline{\textit{Scenario 4 - Two elongated clusters in 3-d - Complete linkage}}} \\
		CH                  & 0.755 & $23*$ & $9$  & $8$  & $4$  & $6$ & $0$ & $0$ & $0$ & $0$ \\
		ASW                 & 1.000 & $50*$ & $0$  & $0$  & $0$  & $0$ & $0$ & $0$ & $0$ & $0$ \\
		Dunn                & 1.000 & $50*$ & $0$  & $0$  & $0$  & $0$ & $0$ & $0$ & $0$ & $0$ \\
		Pearson $\Gamma$        & 0.995 & $49*$ & $1$  & $0$  & $0$  & $0$ & $0$ & $0$ & $0$ & $0$ \\
		Prediction strength  & 0.995 & $49*$ & $1$  & $0$  & $0$  & $0$ & $0$ & $0$ & $0$ & $0$ \\
		Bootstab      & 0.975 & $45*$ & $4$  & $1$  & $0$  & $0$ & $0$ & $0$ & $0$ & $0$ \\
		CVNN      & 0.516 & $1*$  & $6$  & $24$ & $9$  & $7$ & $3$ & $0$ & $0$ & $0$ \\
		${\cal A}_1$& 0.965 & $43*$ & $6$  & $1$ & $ 0$ & $0$ & $0$ & $0$ & $0$ & $0$ \\
		${\cal A}_2$ & 1.000 & $50*$ & $ 0$ & $0$ & $0$ & $0$ & $0$ & $0$ & $0$ & $0$ \\
		
		\hline  
		\multicolumn{11}{c}{\underline{\textit{Scenario 5 - Two ring-shaped clusters in 2-d - Single linkage}}} \\
		CH                  & 0.646 & $0*$  & $4$  & $8$  & $10$ & $11$ & $6$ & $5$  & $3$ & $0$ \\
		ASW                 & 0.711 & $2*$  & $17$ & $13$ & $8$  & $6$  & $2$ & $1$  & $1$ & $0$ \\
		Dunn                & 1.000 & $50*$ & $0$  & $0$  & $0$  & $0$  & $0$ & $0$  & $0$ & $0$ \\
		Pearson $\Gamma$        & 0.617 & $0*$  & $0$  & $0$  & $3$  & $7$  & $4$ & $11$ & $15$ & $0$ \\
		Prediction strength  & 1.000 & $50*$ & $0$  & $0$  & $0$  & $0$  & $0$ & $0$  & $0$ & $0$ \\
		Bootstab      & 0.901 & $37*$ & $0$  & $0$  & $0$  & $0$  & $3$ & $2$  & $5$ & $0$ \\
		CVNN      & 0.736 & $5*$  & $15$ & $16$ & $7$  & $5$  & $2$ & $0$  & $0$ & $0$ \\
		${\cal A}_1$& 0.602 & $ 0*$ & $0$  & $0$ & $ 0$ & $0$ & $2$ & $4$ & $11$ & $33$ \\
		${\cal A}_2$ & 0.982 & $47*$ & $0$ & $0$ & $1$ & $1$  & $0$  & $1$  & $0$  & $0$ \\
		
		\hline  
		\multicolumn{11}{c}{\underline{\textit{Scenario 6 - Two moon-shaped clusters in 2-d - Spectral clustering}}} \\
		CH                  & 0.296 & $0*$  & $0$ & $3$  & $3$  & $8$  & $5$ & $7$ & $11$ & $13$ \\
		ASW                 & 0.349 & $0*$  & $0$ & $6$  & $10$ & $12$ & $6$ & $6$ & $7$  & $3$ \\
		Dunn                & 1.000 & $50*$ & $0$ & $0$  & $0$  & $0$  & $0$ & $0$ & $0$  & $0$ \\
		Pearson $\Gamma$        & 0.452 & $0*$  & $0$ & $14$ & $21$ & $10$ & $2$ & $3$ & $0$  & $0$ \\
		Prediction strength  & 1.000 & $50*$ & $0$ & $0$  & $0$  & $0$  & $0$ & $0$ & $0$  & $0$ \\
		Bootstab      & 0.245 & $0*$  & $0$ & $0$  & $0$  & $0$  & $0$ & $0$ & $6$  & $44$ \\
		CVNN      & 0.338 & $0*$  & $0$ & $5$  & $9$  & $11$ & $7$ & $5$ & $8$  & $5$ \\
		${\cal A}_1$& 0.321 & $ 0*$ & $0$  & $0$ & $ 3$ & $10$ & $9$ & $ 9$ & $11$ & $ 8$ \\
		${\cal A}_2$ & 1.000 & $50*$ & $0$ & $0$ & $0$ & $0$ & $0$ & $0$  & $0$  & $0$ \\
		\thickhline 
		\end{tabular}
	\label{tab:simulated_data_clust_valid_res2}	
\end{table*}

Tables~\ref{tab:simulated_data_clust_valid_res} and \ref{tab:simulated_data_clust_valid_res2} give two kinds of results, namely the distribution of the estimated numbers of clusters, and the average ARI (the maximum ARI is 1 for perfect recovery of the true clusters; a value of 0 is the expected value for comparing two unrelated random clusterings, negative values can occur as well). In case that the number of clusters is estimated wrongly, arguably finding a clustering with high ARI and therefore similar to the true one is more important than having the number of clusters close to the true one, and in general it is not necessarily the case that a ``better'' number of clusters estimate also yields a ``better'' clustering in the sense of higher ARI.   

Scenario 1 was rather easy, with many indexes getting the number of clusters always right. The clusters here are rather compact, and ${\cal A}_1$ is among the best validation methods, with ${\cal A}_2$ lagging somewhat behind. 

In Scenario 2, clusters are still spherical. CVNN does the best job here and finds the correct number of clusters 45 times. Although ${\cal A}_1$ manages this only 36 times, the average ARI with 0.930 is almost the same as what CVNN achieves, both better by some distance than all the other methods. ${\cal A}_2$ once more performs weakly. This should have been a good scenario for CH, because clusters are still spherical Gaussian, but compared to scenario 1 it loses quality considerably, probably due to the higher dimensionality.

In scenario 3, the ARI was computed based on 6 clusters, but 4 clusters are seen as a sensible estimate of $K$. Bootstab achieves the best ARI result, followed closely by ${\cal A}_1$, which gets the number of clusters right most often (and estimated $K=6$ more often than $K=5$ as only method), and ${\cal A}_2$. The other methods are by some distance behind. 

In scenario 4, where elongated clusters mean that some within-cluster distances are quite large, ${\cal A}_2$ performs better than ${\cal A}_1$, which puts more emphasis on within-cluster homogeneity. Apart from ${\cal A}_2$, also ASW and the Dunn index deliver a perfect performance. CH and particularly CVNN are weak here; the other methods are good with the occasional miss. 
  
In Scenario 5, within-cluster homogeneity is no longer a key feature of the clusters. ${\cal A}_2$ does an almost perfect job (Dunn and PS achieve $ARI=1$), whereas ${\cal A}_1$ is much worse, as are CH and Pearson$\Gamma$, with ASW and CVNN somewhat but not much better.

Scenario 6 produces similar results to Scenario 5 with once more ${\cal A}_2$, Dunn and PS performing flawlessly. The rest is much worse, with Pearson$\Gamma$ here best of the weaker methods and Bootstab in last position.

Overall these simulations demonstrate convincingly that different clustering problems require different cluster characteristics, as are measured by different indexes. One of ${\cal A}_1$ and ${\cal A}_2$ was always among the best methods, depending on whether the scenario was characterised by a high degree of within-cluster homogeneity, in which case ${\cal A}_1$ did well, whereas ${\cal A}_2$ was the best method where between-clusters separation dominated, and for nonlinear clusters. The results also show that no method is universally good. ${\cal A}_1$ performed very weakly in Scenarios 5 and 6,  ${\cal A}_2$ failed particularly in scenario 2, and ended up in some distance to the best methods in scenario 1 and 3. CH was weak in scenarios 3, 5, and 6 and suboptimal elsewhere, ASW failed in scenarios 3 and 6, and was suboptimal in some others, the Dunn index did not perform well in scenarios 1-3, Pearson$\Gamma$ was behind in scenarios 2, 3, 5, and 6, PS failed in scenarios 2 and 3, Bootstab in Scenarios 2 and 6, and CVNN in Scenarios 3-6. In any case, when splitting up the scenarios into two groups, namely scenario 1, 2, and 4, where homogeneity and dissimilarity representation are more important, and 3, 5, and 6, where separation is more important, ${\cal A}_1$ on average is clearly the best in the first group with an average ARI of 0.962, with Pearson$\Gamma$ achieving 0.928 in second place, and ${\cal A}_2$ is clearly the best in the second group with an average ARI of 0.907 followed by Dunn achieving 0.857. The assignment of scenarios 3 and 4 may be controversial. If scenario 3 is assigned to the homogeneity group, ${\cal A}_1$ is best by an even larger margin. If in exchange scenario 4 is assigned to the separation group, Dunn, PS, and  ${\cal A}_2$ all achieve an average ARI better than 0.99.  
${\cal A}_2$ also has the best overall mean of 0.895, although this is less relevant because all overall means are affected by bad results in at least some scenarios, and no method should be recommended for universal use (which was ignored in almost all introductory papers of the already existing methods).     

The composite indexes ${\cal A}_1$ and ${\cal A}_2$ have a clear interpretation in terms of the features that a good clustering should have, and the results show that they perform in line with this interpretation. The researcher needs to decide what characteristics are required, and if this is decided correctly, a good result can be achieved. Obviously in reality the researcher does not know, without having clustered the data already, what features the ``true'' clusters have. However, in many real applications there is either no such thing as ``true clusters'', or the situation is ambiguous, and depending on the cluster concept several different clusterings could count as ``true'', see \cite{hennig15true}. In a certain sense, by choosing the cluster concept of interest, the researcher ``defines'' the ``true clusters''.

\subsection{Real data examples with given classes}

In this section we analyse 
three data sets obtained from the University of California Irvine Machine Learning Repository (\cite{Dua2017}) with given classes. Following the approach motivated above, we do not make heavy use of subject-matter information here in order to decide which indexes to aggregate. We list the three best clusterings nominated by the different indexes. In real data analysis it it recommended to not only consider the ``optimum'' solution, because it can be very informative to know that some other potentially quite similar clusterings are similarly good in terms of the used validity index. We also show some exemplary plots that compare all clustering solutions. In Figures \ref{fwinedata}, \ref{fseedsdata}, and \ref{fmovementdata}, discriminant coordinates (DC; \cite{seber1984}) are shown, which optimise the aggregated squared distances between cluster means standardised by the pooled within-cluster variances. 

Out of the composite indexes ${\cal A}_1$ and ${\cal A}_2$, ${\cal A}_1$ comes out very well compared to the other indexes, whereas ${\cal A}_2$ picks among its three best clusterings only single linkage solutions isolating outlying points, achieving ARI values around zero (results not shown). This indicates that for many real data sets with lots of random variation, separation with flexibility in cluster shapes and potentially large within-cluster distances is not a characteristic that will produce convincing clusters. Many meaningful subpopulations in real data are not strongly separated, but come with outliers, and separation-based indexes have then a tendency to declare well separated outliers as clusters.

\subsubsection{Wine data}


This data set is based on the results of a chemical analysis of three types of wine grown in the same region in Italy. The Wine data set was first investigated in \citet{forina1988parvus}. It contains 13 continuous variables and a class variable with 3 classes of sizes $48$, $59$, and $71$. 

\begin{figure*}[tb]
\begin{center}
\includegraphics[width=0.485\textwidth]{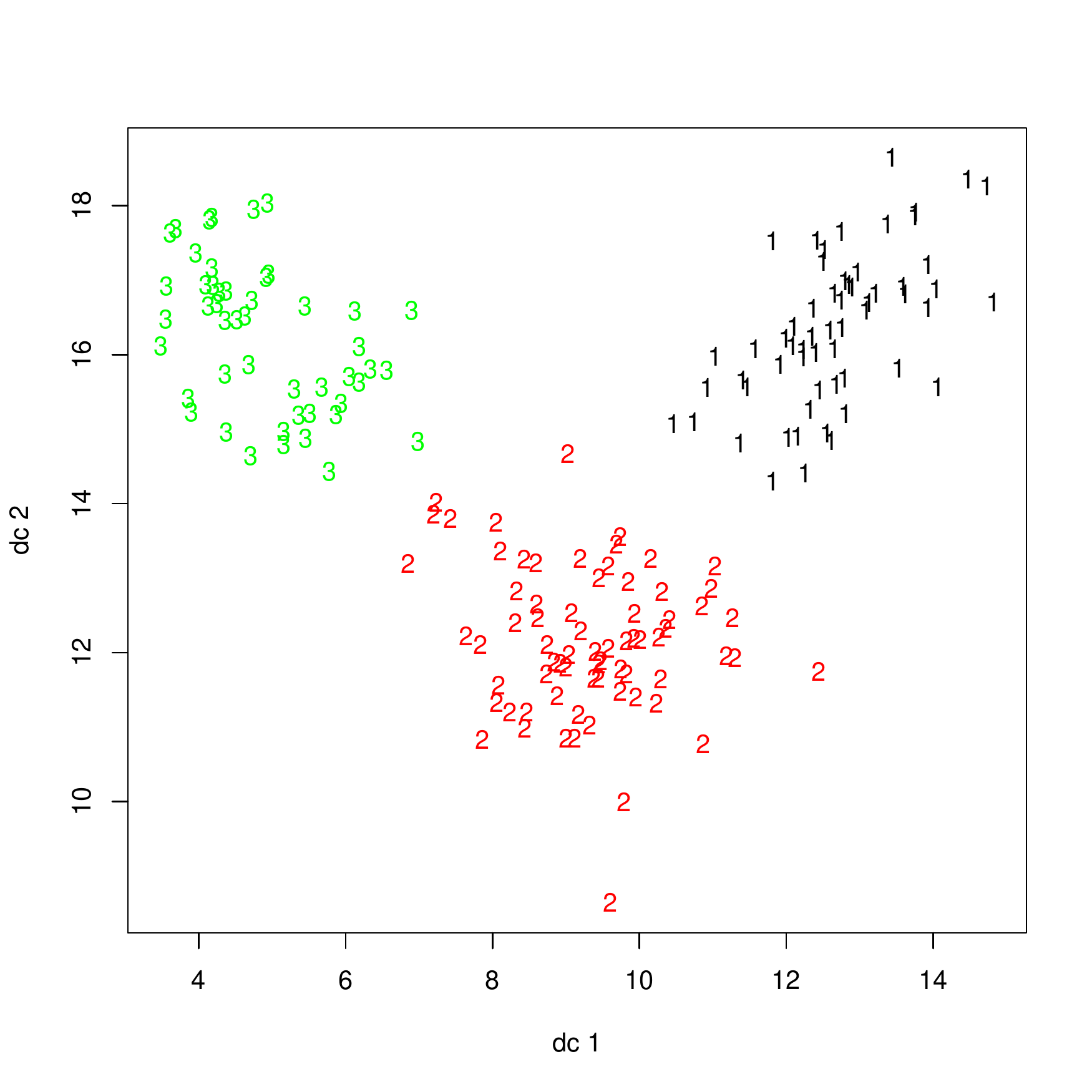}
\includegraphics[width=0.485\textwidth]{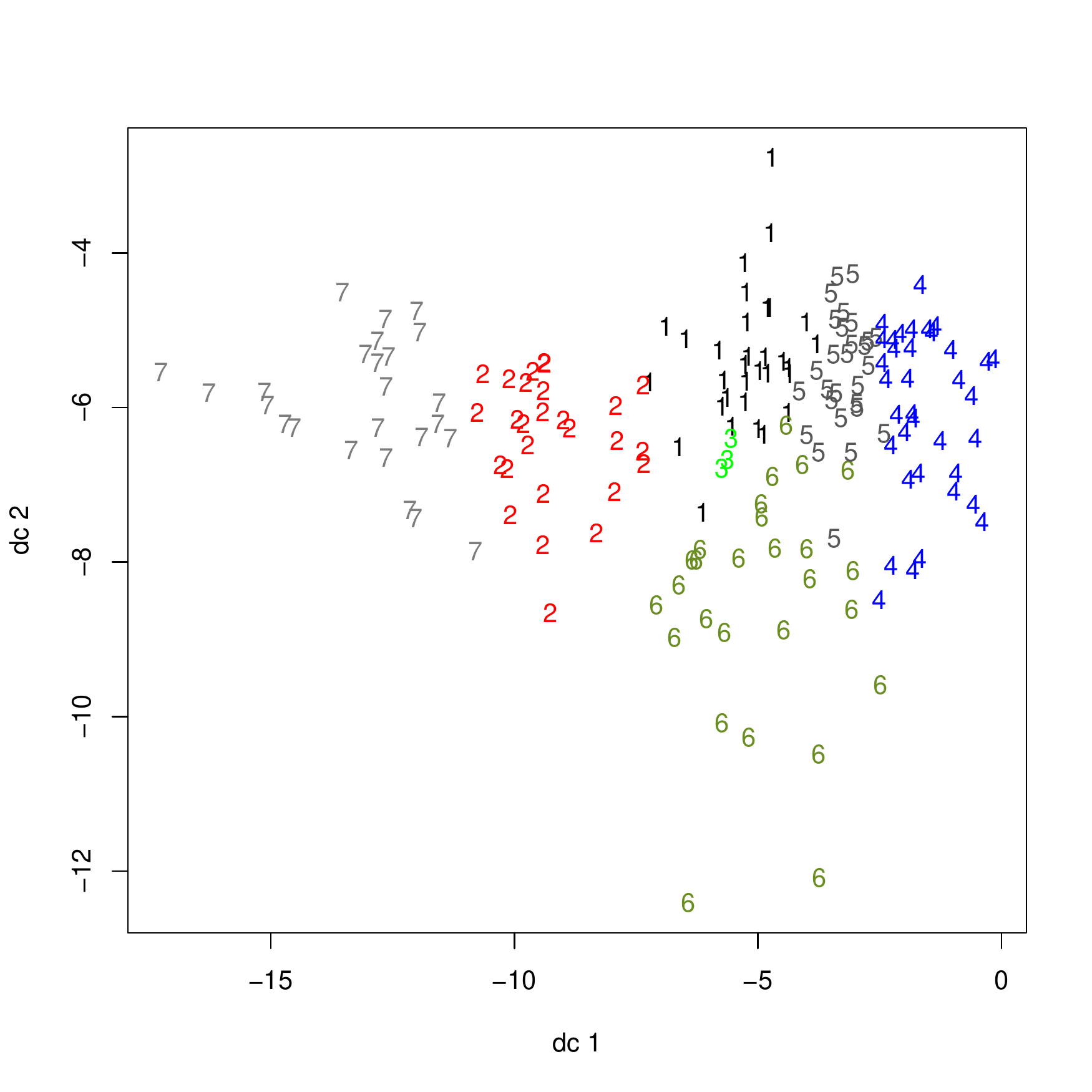}
\end{center}
\caption{Discriminant coordinates plots for Wine data. Left side: True classes. Right side: Clustering solution by spectral clustering with K = 7 (DCs are computed separately for the different clusterings).}
\label{fwinedata}
\end{figure*}

\begin{figure*}[tb]
\begin{center}
\includegraphics[width=0.485\textwidth]{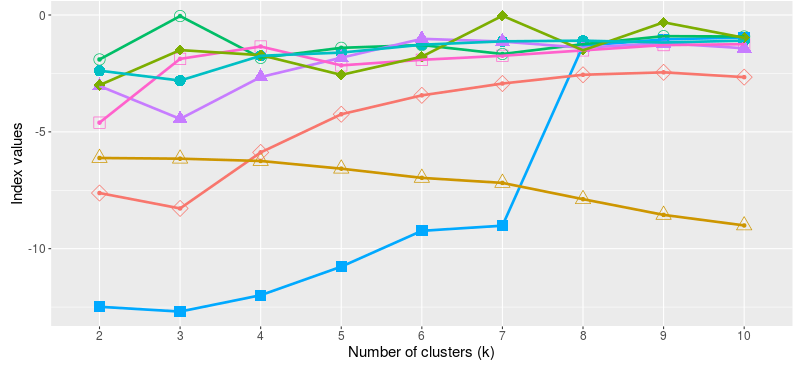}
\includegraphics[width=0.485\textwidth]{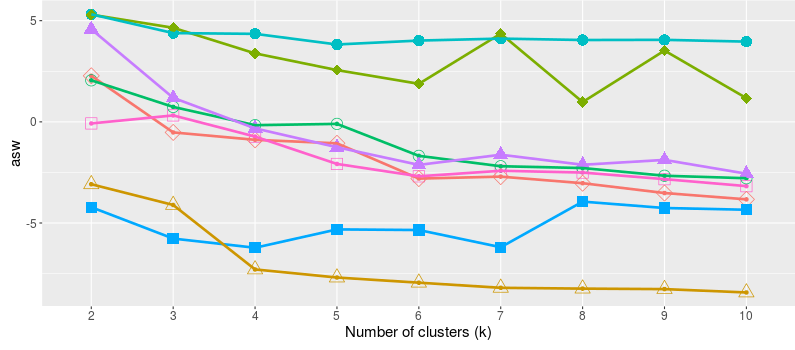}
\includegraphics[width=0.485\textwidth]{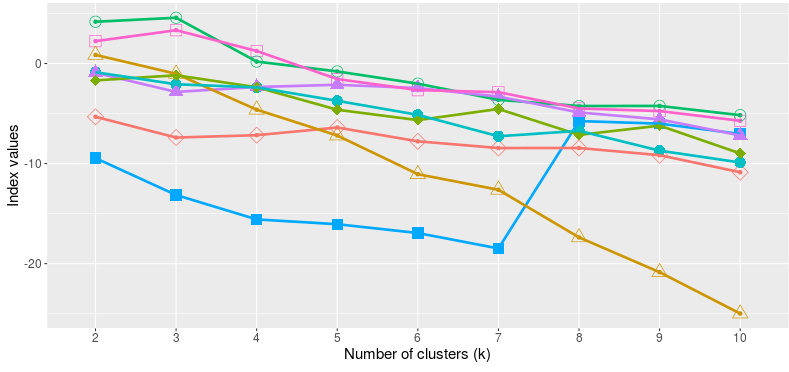}
\includegraphics[width=0.485\textwidth]{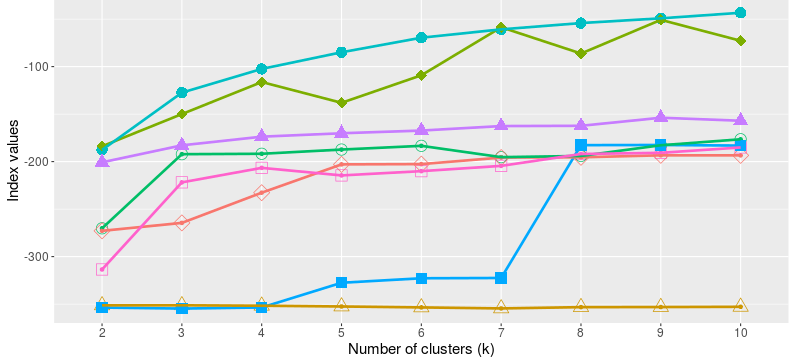}
\includegraphics[width=0.35\textwidth]{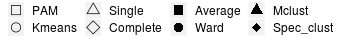}
\includegraphics[width=0.35\textwidth]{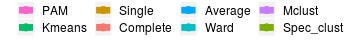}
\end{center}
\caption{Results for Wine data. First row left: ${\cal A}_1$ index. 
First row right: ASW index ($Z$-score calibrated). Second row left:  ${\cal A}_1$ index with calibration based on same $K$ only. Second row right: ${\cal A}_1$ index based on aggregation without calibration.}
\label{fwineresults1}
\end{figure*}

\begin{figure*}[tb]
\begin{center}
\includegraphics[width=0.485\textwidth]{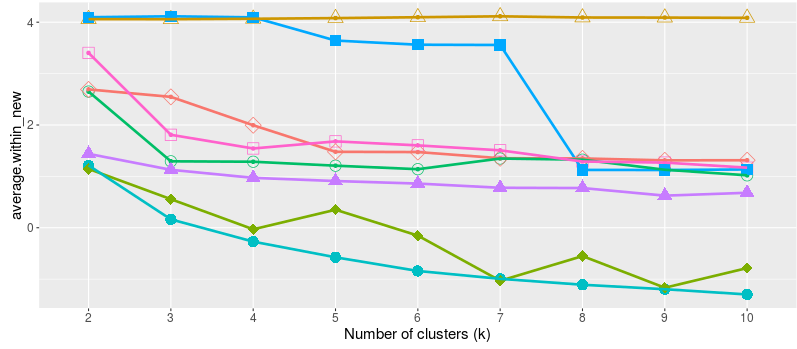}
\includegraphics[width=0.485\textwidth]{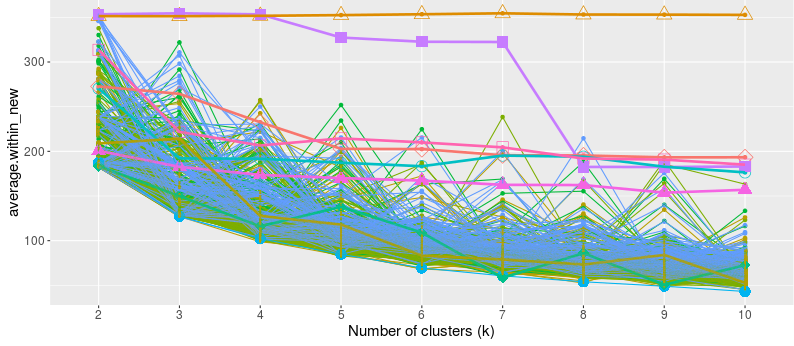}
\includegraphics[width=0.485\textwidth]{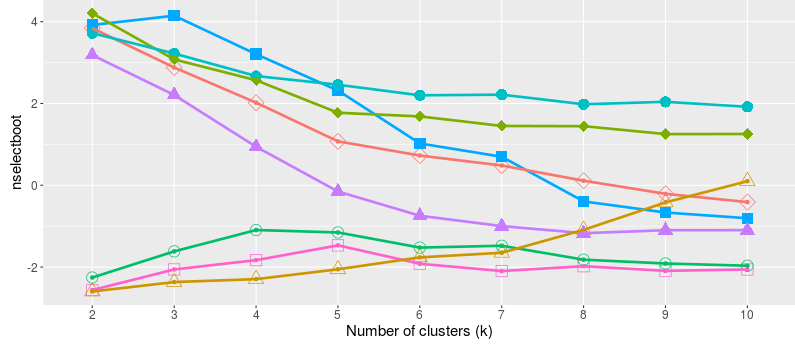}
\includegraphics[width=0.485\textwidth]{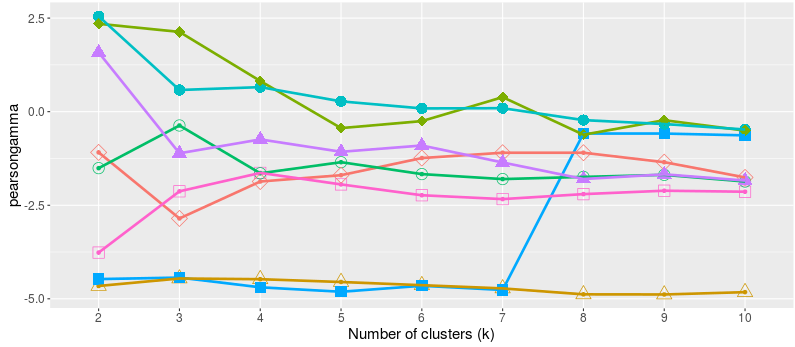}
\includegraphics[width=0.35\textwidth]{bwsymbols.png}
\includegraphics[width=0.35\textwidth]{colorlines.png}
\end{center}
\caption{Results for Wine data. First row left: $I_{ave.wit}$ ($Z$-score calibrated).
First row right: $I_{ave.wit}$ without calibration, with random clusterings. Second row left: Bootstab ($Z$-score calibrated). Second row right:  Pearson$\Gamma$ ($Z$-score calibrated). }
\label{fwineresults2}
\end{figure*}

Figure \ref{fwinedata} (left side) shows that a two-dimensional projection of the data can be found where the different wine types are well separated, but this is difficult to find for any clustering method, particularly because the dimension is fairly high given the low number of observations. 
The right side shows the DC plot for the 7-cluster solution found by spectral clustering, which is the best according to ${\cal A}_1$. 
ARI results comparing the clusterings with the true grouping are given in Table \ref{tab:real_data_ari_comparison}. The 3-means solution is the best, achieving an ARI of 0.9. According to ${\cal A}_1$ this is second best. ${\cal A}_1$ is the only validity index that chooses this clustering among its top three. This also makes ${\cal A}_1$ the best index regarding the average ARI over the top three; the next best clustering picked by any of the indexes has an ARI of 0.37. 

Figures \ref{fwineresults1} and \ref{fwineresults2} show results. Figure \ref{fwineresults1} shows the complete results for ${\cal A}_1$. Added in the top row are results for ASW ($Z$-score calibrated), selected for reasons of illustration. One thing to note is that the best values of ${\cal A}_1$ are not much above 0, meaning that they are on average not much better than the results for the random clusterings. In fact, for larger values of $K$, the random clusterings produce better results than all the proper clustering methods. Looking at the lower left side plot, in which results are shown calibrated separately for separate $K$, it can be seen that for two and three clusters the best proper clustering methods (which are also the best overall, so the 3-means solution with ARI 0.9 comes out as the best here) are still superior to the random clusterings, meaning that calibrated index values are substantially larger than zero. For $K\ge 5$ they all drop below zero. A possible explanation is that the proper clustering methods for larger $K$ fail by trying to adapt to a structure that does not exist, whereas the random clusterings are less affected by the fact that their $K$ does not correspond to a meaningful number of clusters. Looking at more detail in Figure \ref{fwineresults2}, it can be seen that even for larger $K$ the random clusterings are beaten by the best proper clustering methods regarding $I_{ave.wit}$ (the first row shows the $Z$-calibrated values and the raw values with random clusterings added; note that in these plots as well as for Bootstab smaller values are better, as opposed to ${\cal A}_1$, Pearson$\Gamma$, and ASW), and also for Bootstab (lower left plot in Figure \ref{fwineresults2}). Regarding Pearson$\Gamma$ (lower right plot in Figure \ref{fwineresults2}), the best proper clusterings are slightly below the average of the random clusterings for larger $K$ ($Z$-calibrated values below 0), but different methods are best for different criteria, and the random clusterings are better for large $K$ after aggregation. Figure \ref{fwineresults1} (upper right) shows that for the ASW only the two best methods, Ward and spectral clustering, have $Z$-calibrated values larger than zero for $K\ge 5$, so the issue does not only affect ${\cal A}_1$ and its constituents. Furthermore, ASW suggests the smallest possible $K=2$ as optimal, which can be observed in many applications (e.g., \cite{hennig2013find}) and may benefit from calibration for separate $K$.  

The lower right plot in Figure \ref{fwineresults1} shows that if the indexes are aggregated without calibration, the result is entirely dominated by $I_{ave.wit}$, which in uncalibrated state has the by far largest value range and variance (compare the upper plots in Figure \ref{fwineresults2}, taking into account the sign change for aggregation). This is not sensible.  

\subsubsection{Seeds data}

The Seeds data set (\cite{charytanowicz2010complete}) has measurements of geometrical properties of kernels belonging to three different varieties of wheat. It contains seven geometric features of wheat kernels and a class variable indicating three classes containing 50 objects each. Each class refers to a type of wheat. 

\begin{figure*}[tb]
\begin{center}
\includegraphics[width=0.485\textwidth]{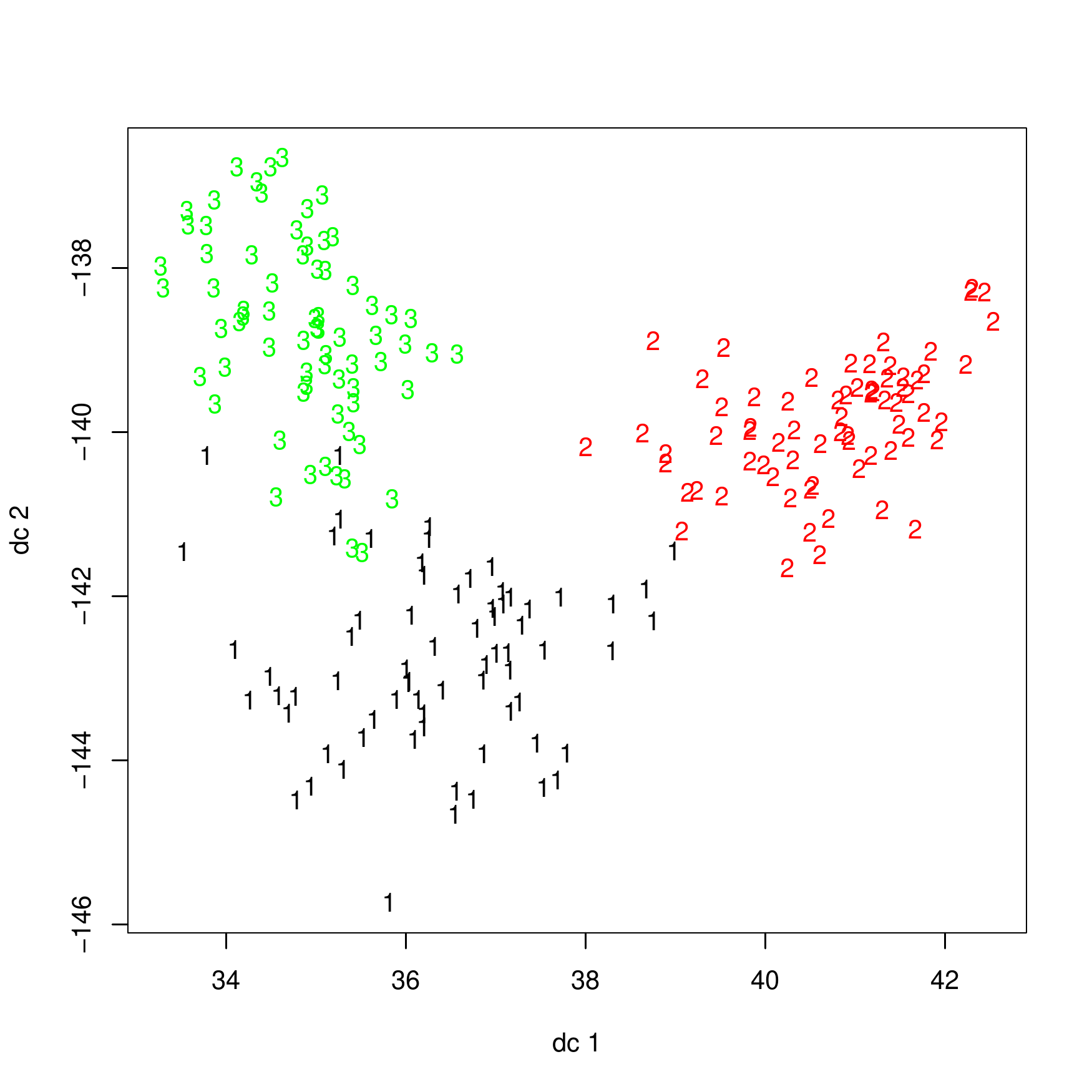}
\includegraphics[width=0.485\textwidth]{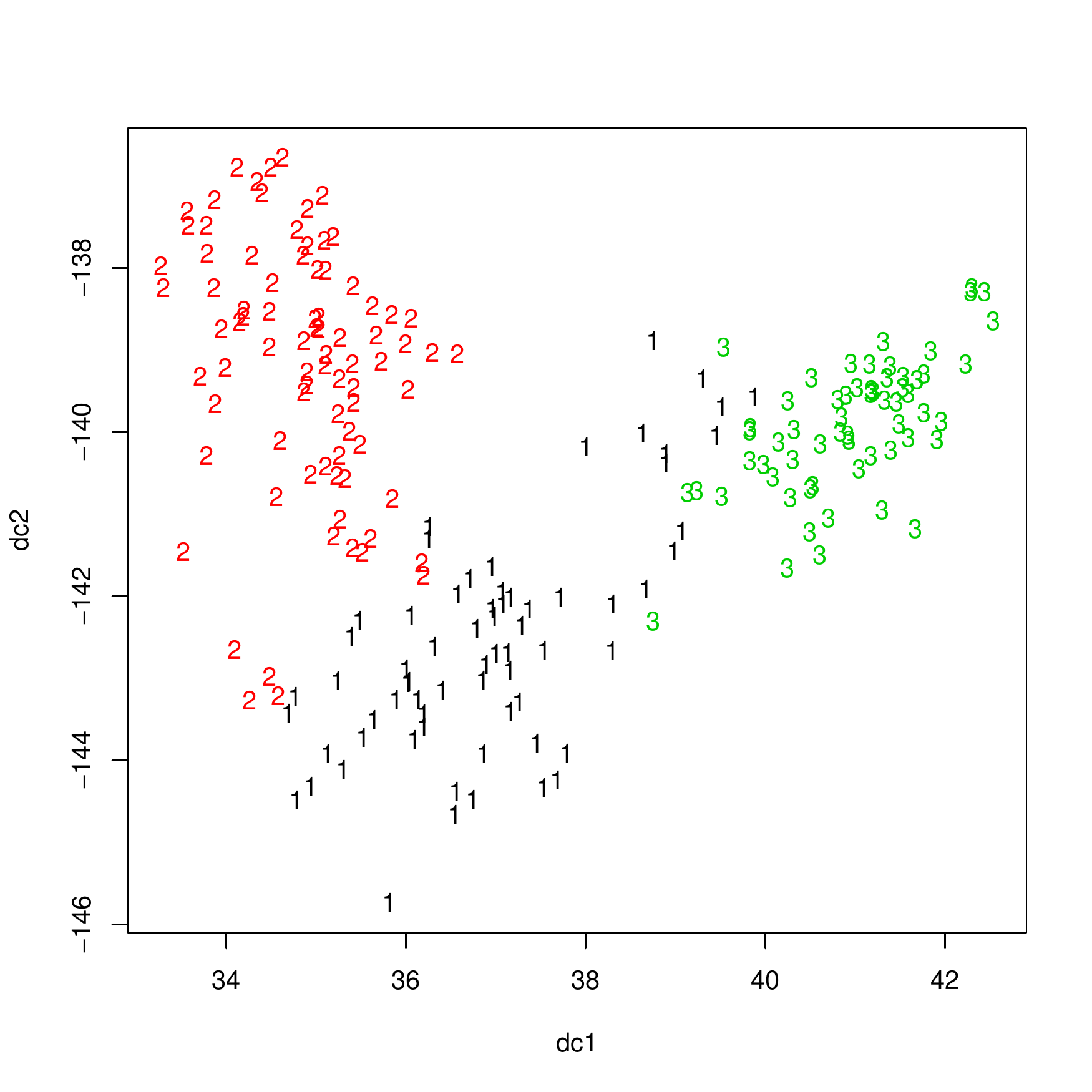}
\end{center}
\caption{Discriminant coordinates plot for Seeds data. Left side: True classes.
Right side: Clustering solution by PAM with $K=3$.}
\label{fseedsdata}
\end{figure*}

\begin{figure*}[tb]
\begin{center}
\includegraphics[width=0.485\textwidth]{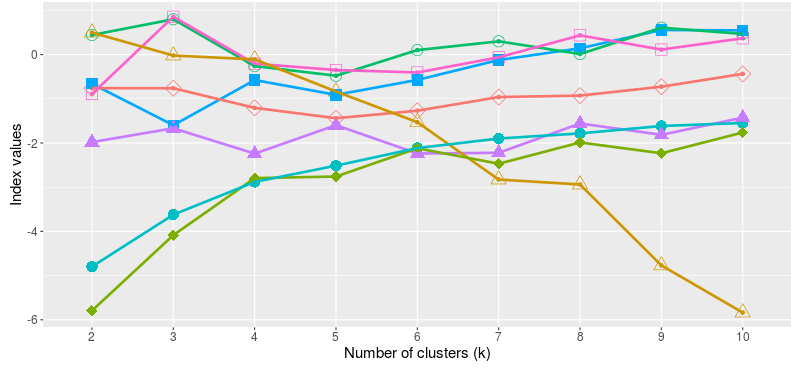}
\includegraphics[width=0.485\textwidth]{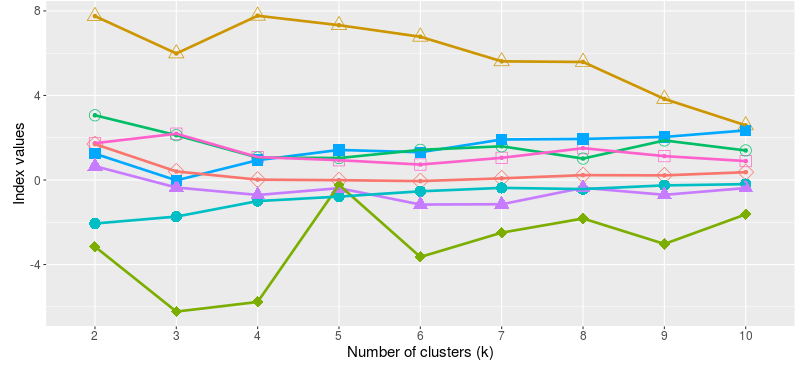}
\includegraphics[width=0.485\textwidth]{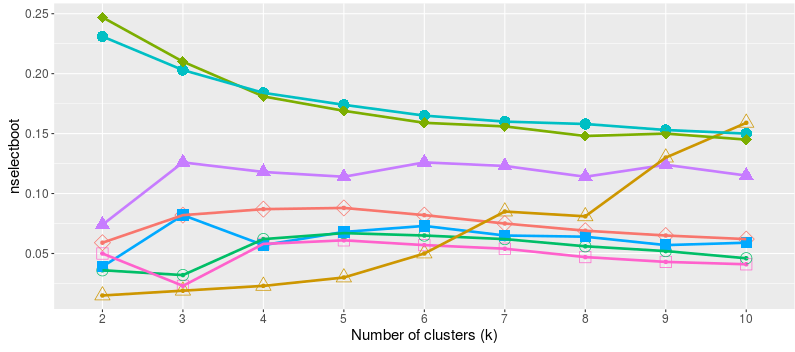}
\includegraphics[width=0.485\textwidth]{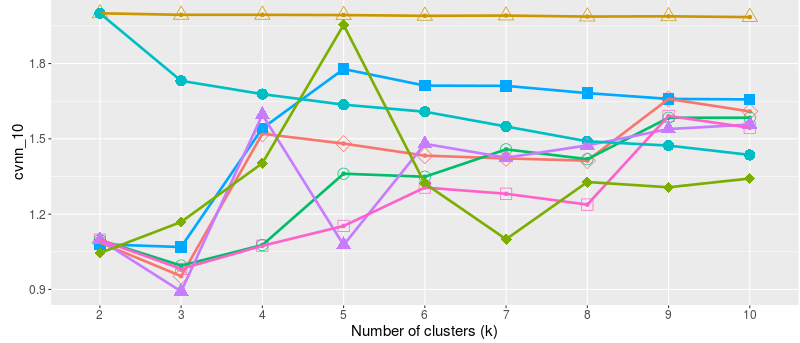}
\includegraphics[width=0.35\textwidth]{bwsymbols.png}
\includegraphics[width=0.35\textwidth]{colorlines.png}
\end{center}
\caption{Results for Seeds data. First row left: ${\cal A}_1$. 
First row right: ${\cal A}_2$. Second row left: Bootstab (uncalibrated; smaller values are better). Second row right: CVNN (uncalibrated; smaller values are better).}
\label{fseedsresults}
\end{figure*}

Figure \ref{fseedsdata} shows DCs for the true three classes (upper left) and the PAM-clustering with $K=3$ (upper right) on the same DCs. This is the clustering that was picked as the best by ${\cal A}_1$, and has the second best ARI (0.75) of all clusterings; only 3-means is slightly better (0.77).

Results are shown in Table \ref{tab:real_data_ari_comparison}. ${\cal A}_1$ picks the two best clusterings as best two, if in reverse order. The clusterings picked as best by all other indexes are worse, however CH and CVNN achieve slightly better ARI averages over the best three clusterings because of better clusterings in third place. The results of ASW, Dunn, Pearson$\Gamma$, PS and Bootstab are clearly weaker. 

Figure \ref{fseedsresults} shows the full results for ${\cal A}_1$, ${\cal A}_2$, Bootstab and CVNN (the latter two uncalibrated, and smaller values are better). ${\cal A}_2$ is once more dominated by single linkage solutions that isolate outliers. The Bootstab plot shows that spectral clustering and Ward with small $K$ are substantially less stable than the other methods; Bootstab's comparison of methods is somewhat similar between Seeds and Wine data set. The CVNN plot gives a good indication for 3 clusters for a number of clustering methods.  

\begin{table*}[tb]
	\renewcommand{\arraystretch}{1.35}
	\tiny
	\caption{Results of real data examples.}
	\centering
		\begin{tabular}{l | ccc|c | rrr}
		\multirow{2}{*}{\textbf{Validity Index}} & \multicolumn{3}{c|}{\textbf{\underline{ARI}}} & \textbf{Average} &  \multicolumn{3}{c}{\textbf{\underline{Best clusterings in order ($K$)}}}\\
		& \textbf{First} & \textbf{Second} & \textbf{Third}& \textbf{ARI} & \textbf{First} & \textbf{Second} & \textbf{Third} \\
		\hline
%
%
		\multicolumn{8}{c}{\underline{\textit{WINE data set}}} \\ 
		CH  & 0.17  & 0.19 & 0.22  & 0.19 & Ward ($10$)  & Ward (9) & Spectral ($7$) \\
		ASW   & 0.33  & 0.37 & 0.31 & 0.34 & Ward ($2$) & Spectral ($2$)  & Spectral ($3$)   \\
		Dunn  & 0.33  & 0.19 & 0.18 & 0.23 & Ward ($2$) & Ward (9) & Spectral ($9$)    \\
		Pearson $\Gamma$  & 0.33 & 0.37 & 0.31 & 0.34 & Ward ($2$) & Spectral ($2$) & Spectral ($3$)  \\
		PS  & -0.01 & -0.01 & -0.01 & -0.01 & Single ($3$) & Single (2) & Single ($4$) \\
		Bootstab & -0.01 & 0.36 & -0.01 & 0.11& Single ($2$) & PAM ($2$) & Single (3) \\
		CVNN  & 0.31 & 0.25 & 0.31 & 0.29 & Spectral ($4$) & Spectral (6) & Spectral (3)\\
		${\cal A}_1$& 0.22 & 0.90 & 0.18 & 0.43 & Spectral (7) & 3-means & Spectral (9) \\
		
		\hline
		\multicolumn{8}{c}{\underline{\textit{SEED data set}}} \\ 		
		CH   & 0.63 & 0.71 & 0.77  & 0.70 & Mclust ($3$)  & Ward ($3$) & $3$-means \\
		ASW  & 0.44  & 0.48 & 0.50 & 0.48 & PAM    ($2$)  & $2$-means & Spectral ($2$) \\
		Dunn   & 0.44  & 0.43 & 0.77  & 0.55 & Ward  ($8$)  & Spectral ($5$) & $3$-means \\
		Pearson $\Gamma$  & 0.63  & 0.44 & 0.71  & 0.60 & Mclust ($3$)  & PAM  ($2$) & Ward ($3$) \\
		PS  & 0.00  & 0.49 & 0.48  & 0.32 & Single ($2$) & Average ($2$) & $2$-means \\
		Bootstab & 0.00 & 0.75 & 0.77 & 0.51 & Single ($2$)  & PAM ($3$) & $3$-means \\
		CVNN & 0.63 & 0.69 & 0.75  & 0.69 & Mclust ($3$) & Complete  ($3$) & PAM ($3$) \\
		${\cal A}_1$& 0.75 & 0.77 & 0.50 & 0.67 & PAM ($3$) & $3$-means & Average ($9$)\\
		\hline
		\multicolumn{8}{c}{\underline{\textit{MOVEMENT data set}}} \\ 
		CH  & 0.07  & 0.05 & 0.04  & 0.05 & 2-means  & Spectral ($2$)  & Average ($2$) \\
		ASW   & 0.30  & 0.31 & 0.24 & 0.28 & 20-means & 19-means  & Ward (10)   \\
		Dunn  & 0.00  & 0.00 & 0.00 & 0.00 & Single ($2$) & Single ($3$)  & Single ($4$)    \\
		Pearson $\Gamma$  & 0.23 & 0.23 & 0.25 & 0.23 & Average ($14$) & Average ($15$) & Average ($16$)  \\
		PS  & 0.00 & 0.00 & 0.00 & 0.00 & Single ($2$) & Single ($3)$ & Single ($4$) \\
		Bootstab & 0.00 & 0.00 & 0.00 & 0.00 & Single ($2$) & Single ($3$) & Single (4) \\
		CVNN  & 0.34 & 0.16 & 0.19 & 0.26 & Spectral ($11$) & Spectral ($4$) & 10-means\\
		${\cal A}_1$& 0.32 & 0.30 & 0.34 & 0.32 & Average (20) & 20-means & Average (18) \\
		\thickhline  
		\end{tabular}
	\label{tab:real_data_ari_comparison}	
\end{table*}

\subsubsection{Movement data} \label{smovement}
The Movement data set \cite{dias2009hand} contains 15 classes of 24 instances each, where each class refers to a hand movement type in LIBRAS, the Brazilian sign language. There are 90 variables, which correspond to 45 positions of two hands tracking hand movement over time extracted from videos. 

Due to the large number of classes and variables together with a still fairly low number of observations, this is a much harder clustering problem than the previous data sets. In this application it is quite clear why within-cluster homogeneity is a more important characteristic than separation, because variation within sign language movement types should be limited (otherwise it would be hard to communicate), whereas strong separation can hardly be expected with a large enough number of movement types. 

\begin{figure*}[tb]
\begin{center}
\includegraphics[width=0.485\textwidth]{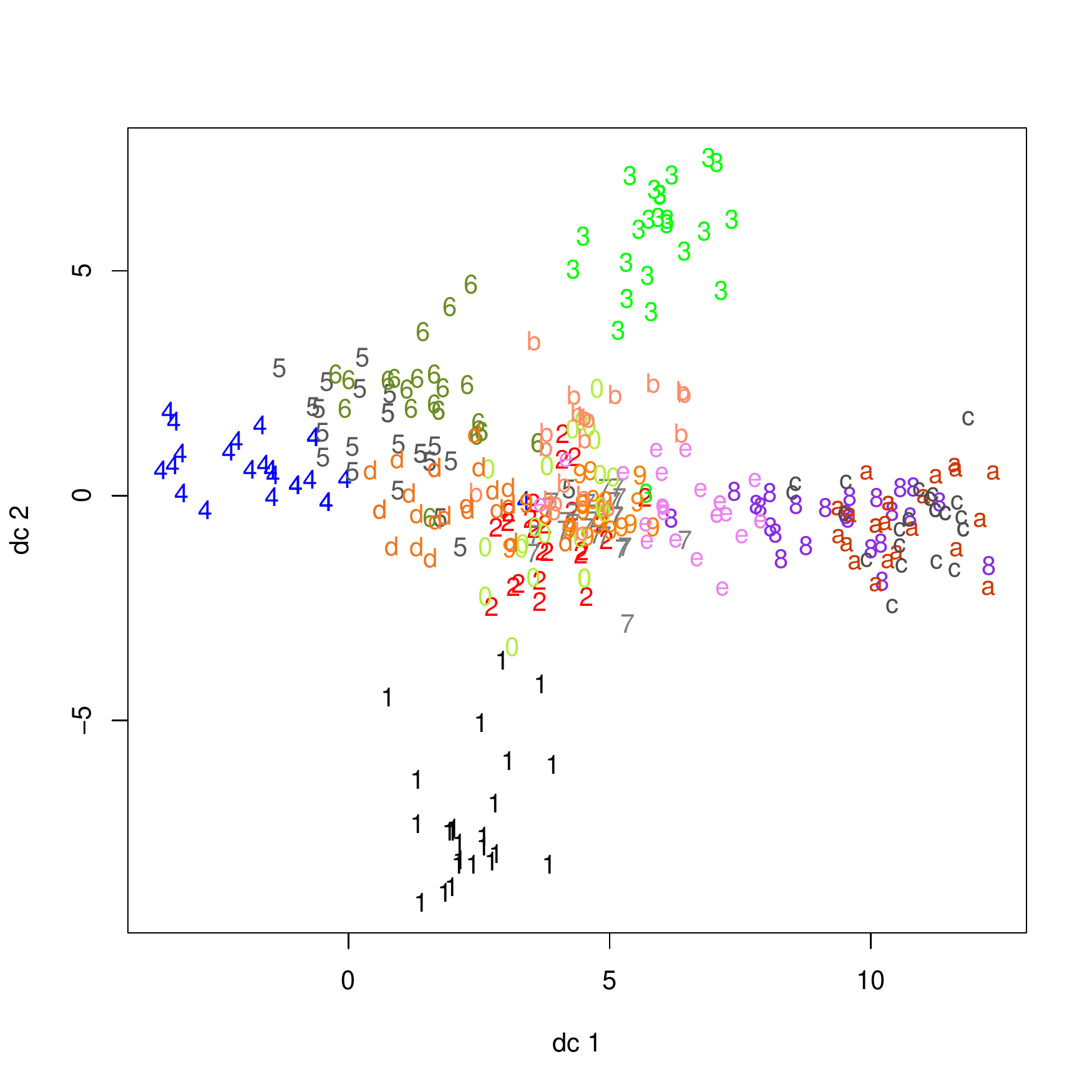}
\includegraphics[width=0.485\textwidth]{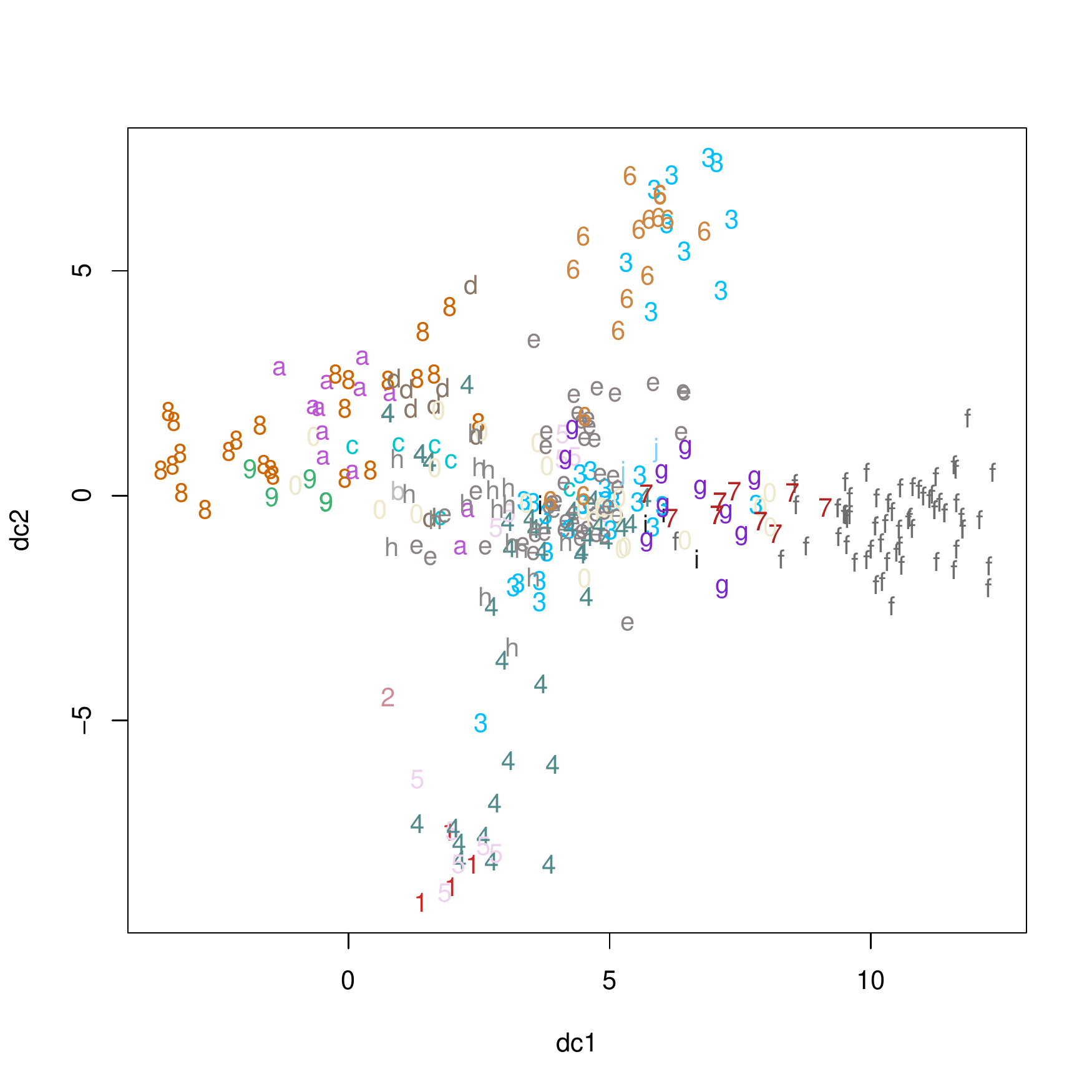}
\end{center}
\caption{Discriminant coordinates plot for Movement data. Left side: True classes.
Right side: Clustering solution by average linkage with $K=20$.}
\label{fmovementdata}
\end{figure*}

\begin{figure*}[tb]
\begin{center}
\includegraphics[width=0.485\textwidth]{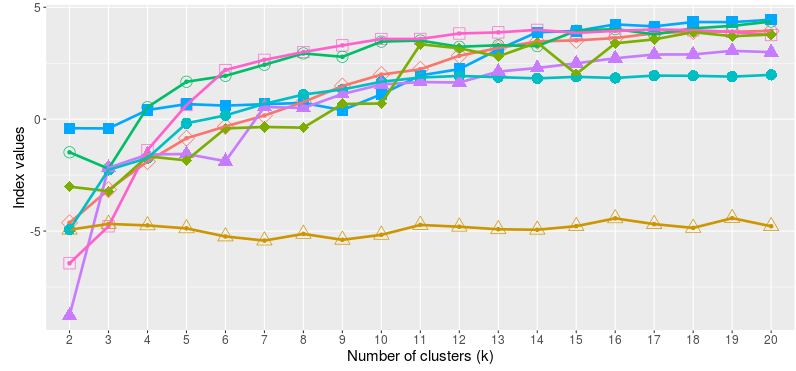}
\includegraphics[width=0.485\textwidth]{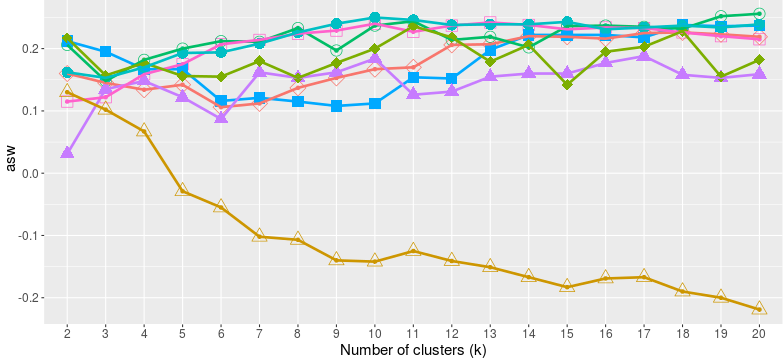}
\includegraphics[width=0.35\textwidth]{bwsymbols.png}
\includegraphics[width=0.35\textwidth]{colorlines.png}
\end{center}
\caption{Results for Movement data. Left: ${\cal A}_1$. Right: ASW (uncalibrated).}
\label{fmovementresults}
\end{figure*}

Figure \ref{fmovementdata} gives the DCs for the true classes on the left side. On the right side the same DCs are used for visualising the $K=20$ solution for average linkage, which is optimal according to ${\cal A}_1$. Obviously the recovery of the true classes is not perfect, but there is clear structure with some connection to the true classes and the DC projection derived from them. The clustering achieves an ARI of 0.32. The largest ARI achieved by any clustering here was 0.39. Out of the validity indexes, CVNN finds one with ARI 0.34 as best, but ${\cal A}_1$ has the best average of the best three clusterings; ASW is the second best according to that metric. For some detailed results see Figure \ref{fmovementresults} and the earlier Figure \ref{fmovementcalibration}. Average linkage turns out to be the best method according to ${\cal A}_1$ for larger numbers of clusters; the index clearly indicates that a larger number of clusters is required here, though not pointing at exactly 15 clusters (which no other index does either; Pearson$\Gamma$ actually delivers the best $K$ around 15, but ARI values of these clusterings are worse). The ASW generates local optima also at $K=2$ and around $K=11$. It prefers $K$-means over average linkage for large $K$, which is slightly worse regarding the ARI. Both methods rule out single linkage clusterings as bad, which is correct here, as opposed to the results of Dunn and  ${\cal A}_2$.

\subsection{German Bundestag data 2005}
The German Bundestag data 2005 is included in the R-package \texttt{flexclust}
(\cite{leisch06}) and 
consists of the German general election 2005 results 
(percentages of second votes, which determine 
the composition of the parliament) 
by the 299 constituencies for the five biggest parties SPD (social democrats, centre left), UNION (Christian conservative, centre right), GRUENE
(green party), FDP (liberal party), and LINKE (left; merger of the  
successor of the communist party from former communist East Germany, 
and a social
justice initiative from the West). Different from the other data examples, this
data set does not come with ``true clusters'' and therefore represents a real 
clustering task, the clustering of constituencies. As in many 
real situations, a fixed and specific aim of clustering is not given. Rather
many uses are conceivable for such a clustering. In particular, clusters can be
used for differentiated analysis of future election results, for simplified 
analysis of complex relationships between voting patterns and other 
characteristic indicators of the constituencies, and for simplified 
election results 
analysis by the media and the political parties themselves. It is therefore 
hard to specify a weighting of required cluster characteristics, although 
it is conceivable to do this for very specific uses. Instead we use the data set
to demonstrate the use of the two basic composite indexes, which we recommend
in absence of information about how to choose the weights. We applied seven
clustering methods (all methods applied earlier except Ward's) with
numbers of clusters between 2 and 12.  

\begin{figure*}[tb]
\begin{center}
\includegraphics[width=0.485\textwidth]{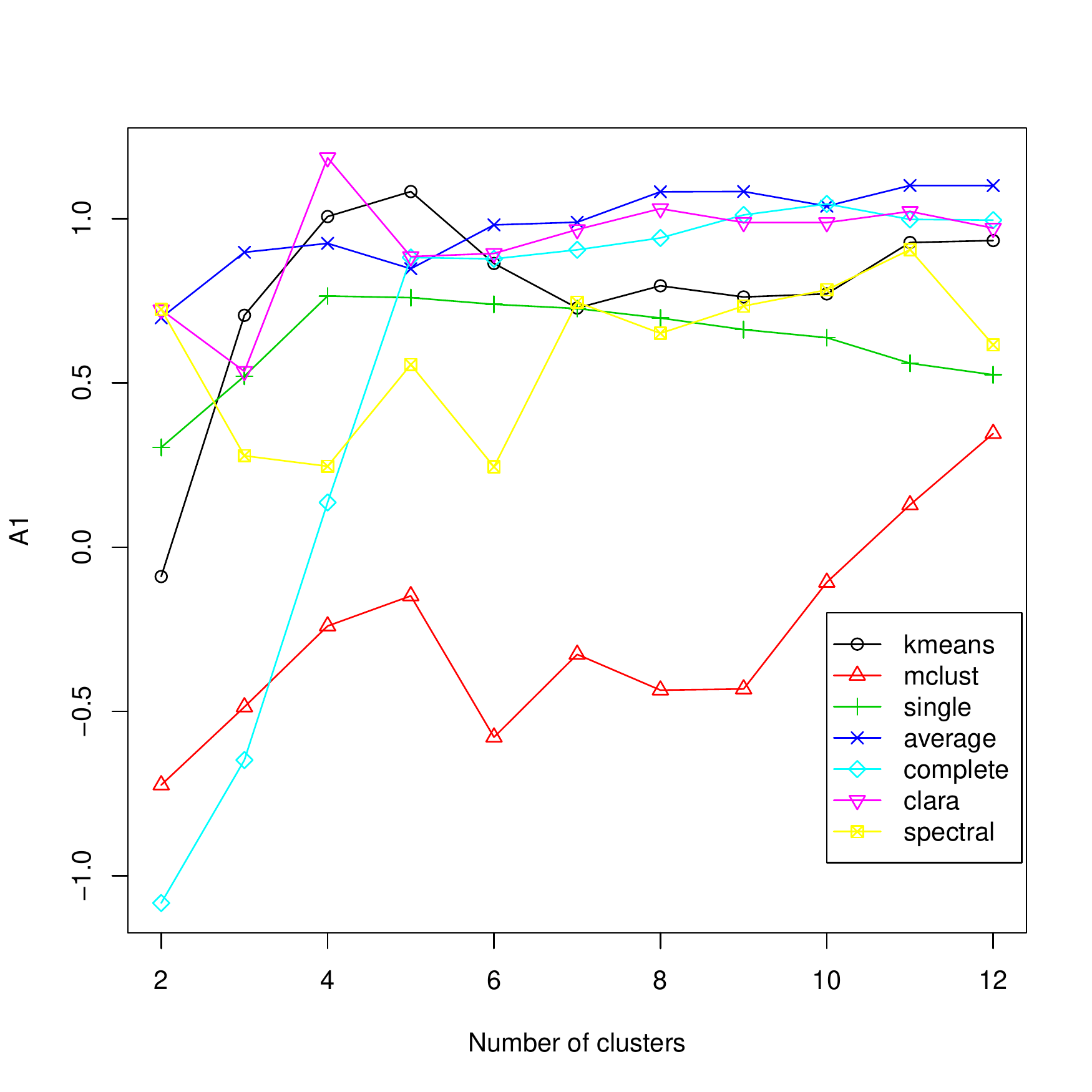}
\includegraphics[width=0.485\textwidth]{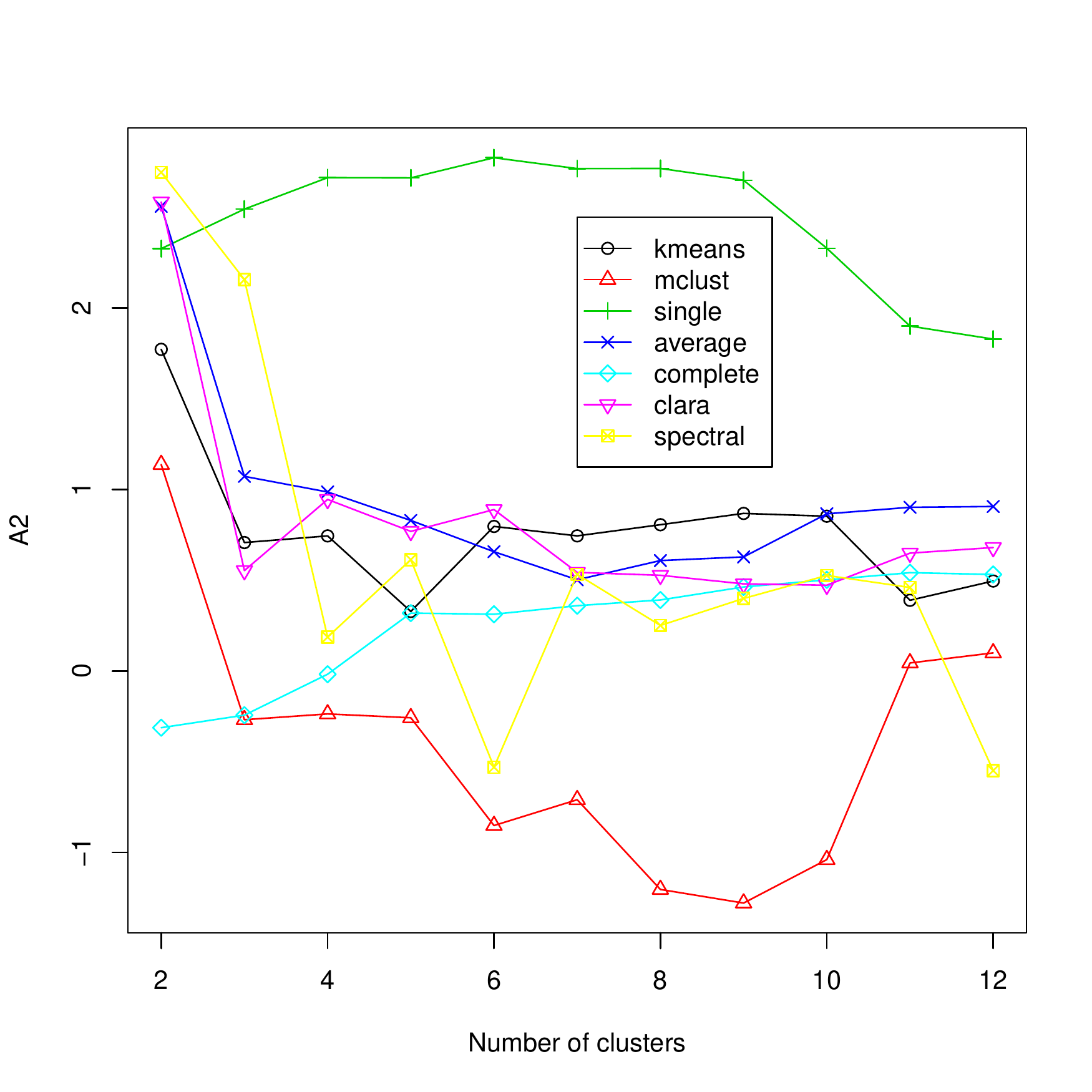}
\end{center}
\caption{Results for Bundestag data. Left: ${\cal A}_1$. 
Right: ${\cal A}_2$.}
\label{fbundestagresults}
\end{figure*}

\begin{figure*}[tb]
\begin{center}
\includegraphics[width=0.485\textwidth]{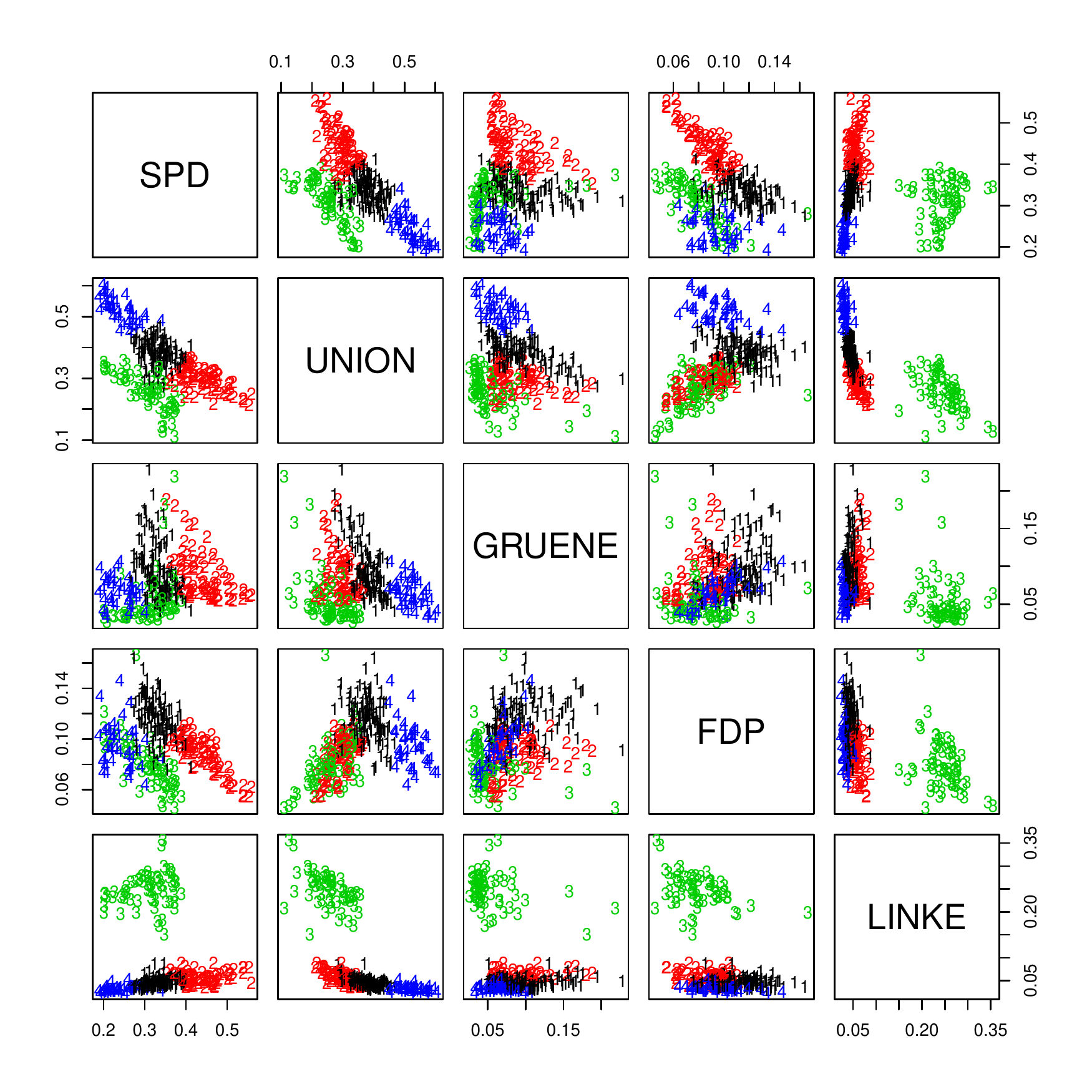}
\includegraphics[width=0.485\textwidth]{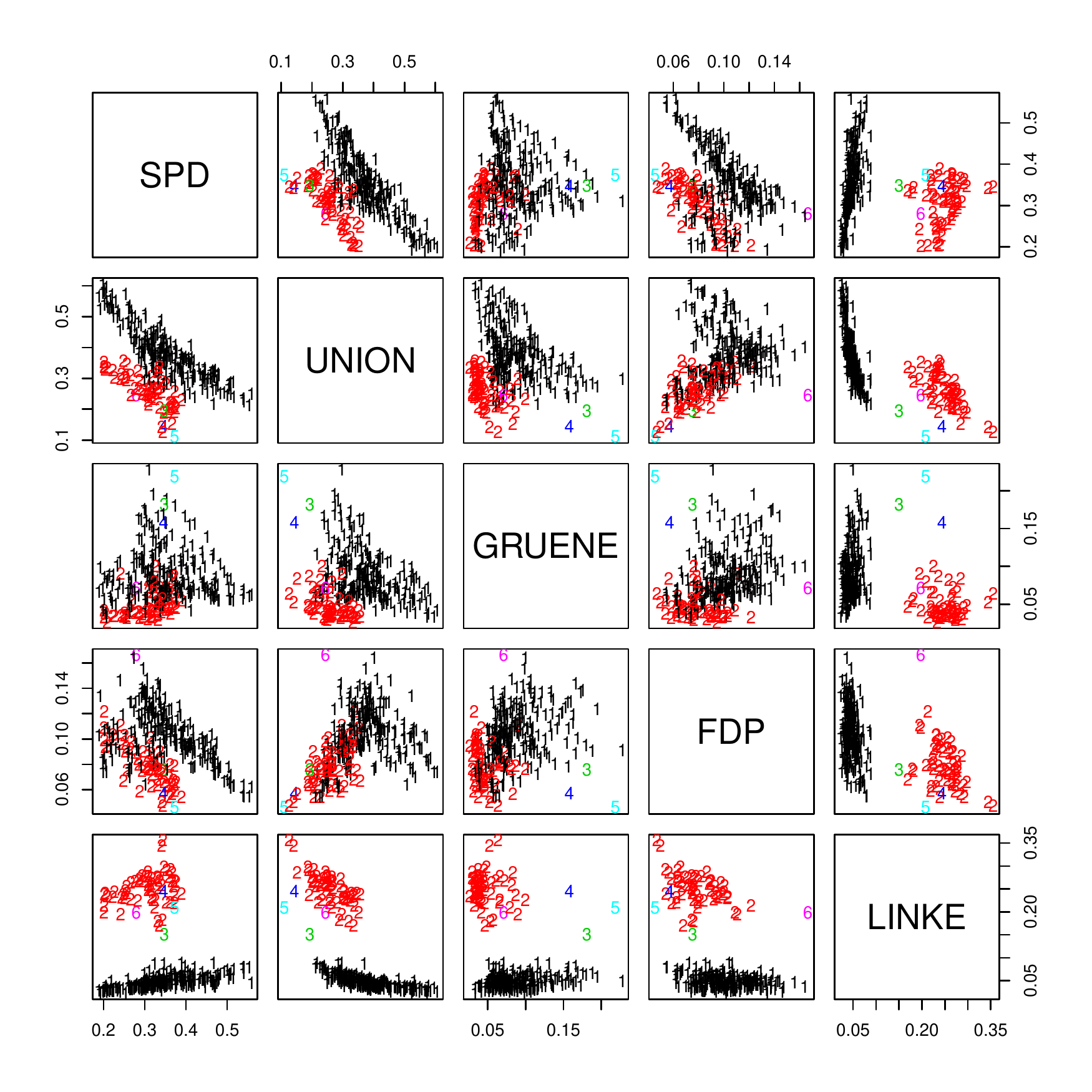}
\end{center}
\caption{Clusterings of Bundestag data. Left: PAM clustering with $K=4$ (${\cal A}_1$-optimal), right: Single Linkage clustering with $K=6$ (${\cal A}_2$-optimal).}
\label{fbundestagcluster}
\end{figure*}

The results are shown in Figure \ref{fbundestagresults}. ${\cal A}_1$ and 
${\cal A}_2$ deliver quite different optimal clusterings, both of which make
sense in different ways. The best clustering according to  ${\cal A}_1$ is from
PAM with $K=4$. The best clustering according to  ${\cal A}_2$ is from Single
Linkage with $K=6$. The data with both clusterings are shown in Figure
\ref{fbundestagcluster}. 

The most obvious feature of the data is the separation into two clearly visible clusters according to the results of the LINKE party. The LINKE are strong in all constituencies in former East Germany plus the four constituencies of the Saarland in the West, which is the homeland of the LINKE top candidate 2005 Oskar Lafontaine, who had been a very popular leader of the Saarland between 1985 and 1998. They are much worse in all other western constituencies. The clustering optimal according to the separation-based index ${\cal A}_2$ basically delivers this pattern, except that four outlying constituencies in the East are isolated as their own clusters. Looking at the data, this clearly makes sense, as these four constituencies are clearly special. Three of them (clusters 3, 4, and 5) are in Berlin, with strong results for the GRUENE unlike the rest of the East, and quite diverse performances of the other parties. The last one is in Dresden and is the only constituency in the whole of Germany that has a strong LINKE result together with a strong FDP. In fact, a candidate in that constituency had died shortly before the election, and the constituency had voted two weeks later than the rest of Germany. Subtleties of the German election law (that have been changed in the meantime) meant that the UNION could have lost a seat in parliament had they performed too well here, apparently prompting some of their voters to vote FDP (\url{https://de.wikipedia.org/wiki/Bundestagswahl_2005}, accessed 27 May 2020).   

The next best clusterings according to ${\cal A}_2$ with $K=2$ basically just split the data set up into the East plus Saarland and the remainder of the West, which also makes data analytic sense.

These clusterings may be too crude for some uses, and homogeneity may be seen as more important. The PAM clustering with $K=4$, optimal according to ${\cal A}_1$, separates the East plus Saarland from the West and splits the West up into three clusters, SPD strongholds (no. 2), CDU strongholds (no. 4), and
no. 1 as a group in between with mostly good FDP results. Looking at the German ``L\"{a}nder'' (administrative regions with much autonomy), many of them belong completely to one of these clusters, and some like the two biggest ones Nordrhein-Westfalen and Bayern split up quite interestingly between two of them, separating more conservative catholic regions with smaller population density from bigger cities with stronger SPD results. If a finer grained clustering is required, the next best ${\cal A}_1$-result, Average Linkage with $K=11$ (not shown) splits up the West into four clusters also including a cluster of GRUENE strongholds, one cluster dominated by the Saarland is separated from the East, and some outliers as discussed before are isolated. Probably some more clusterings could be interpreted in a sensible way, but the best clusterings according to ${\cal A}_1$ and ${\cal A}_2$ are certainly data analytically valid and informative.

\section{Discussion}
\label{sec:discussion}

The general idea of this paper is that cluster validity can be measured by composite indexes, which can be made up from indexes measuring various characteristics of a clustering important to a user in a given application. Meaningful aggregation requires calibration, which can be done against a set of random clusterings generated on the data set. A user may not feel comfortable to make detailed decisions about included indexes and weights. For this reason we suggest two specific composite indexes which focus on either within-cluster homogeneity or between-clusters separation. These give substantially different results and have been shown by our experiments to be complementary advantageous in different situations. However, we recommend to take into account background information and the specifics of the situation and the problem at hand in order to define an composite index that can be better adapted to the problem at hand. Graphical displays such as principal components analysis may also help. Generally we have seen that no index can be expected to work universally, and that different cluster characteristics are important in different situations. The present work enables the user to adapt cluster validation to what counts in the specific application and shows that with an appropriate decision about important basic required characteristics of clusters even the proposed basic composite indexes can outperform indexes existing in the literature.

These basic composite indexes are {\it not} recommended for general automatic use, and the user is encouraged to employ all available background knowledge to design their own composite index for a given situation, and to take into account the dynamics of individual indexes of interest before aggregation to decide about one or more optimum clusterings. Such decisions can be made before knowing results, making the result free from the influence of selective perception or even the suspicion that the user just cherry-picks a clustering that confirms their prior opinion. On the other hand (once more reflecting the earlier mentioned tension between objectivity and flexible adaptation to specifics of the situation), results may point the user to some unanticipated issues and may cause them to change ideas. For example, for the Wine data the user can conclude from the difficulty to find any proper clustering with ${\cal A}_1$ substantially larger than 0 that calibration based on separate $K$ can be more suitable here, even without taking the ``true'' clustering (which in reality is not available) into account.

In many situations background information and knowledge of the aim of clustering may not be enough for the user to confidently choose individual indexes and fix weights, and in such situations we recommend the basic composite indexes as starting points for an analysis, keeping in mind at least the rough distinction regarding whether homogeneity and dissimilarity representation on one hand, or separation and avoidance of within-cluster gaps on the other hand, are the focus in the situation at hand.

The present paper can be seen rather as a starting point than as an end point for research. The simulations and examples are rather exemplary than comprehensive. Among various things that could be tried out, non-Euclidean data using other distances could be a worthwhile topic of investigation. Additional indexes and random clustering methods are conceivable (\cite{hennig2017cluster} proposes some  more). There are various possibilities for theoretical investigation of the indexes and aggregation strategies. An issue with the approach is numerical complexity. Whereas the random clusterings can be quickly generated, computing all index values for all proper and random clusterings takes time. In particular, the stability indexes PS and Bootstab involve resampling, and are heavy to run for a large number of involved clusterings. Shortcuts and investigations how small the number of random clusterings and resamplings can be chosen while still leaving results reasonably stable would be worthwhile (we believe and some experience shows, that good results can already be achieved for, say, $B=20$ and $A=25$), although with the nowadays available computing power and parallel computing data sets clearly larger than those treated in this paper (although probably not with millions of observations and thousands of variables) can be analysed. 

The methodology presented here is implemented in function ``clusterbenchstats'' in the R-package ``fpc''.

{\bf Acknowledgements:}
The work of the second author was supported by EPSRC grant EP/K033972/1.

%
%

\bibliographystyle{chicago}

\bibliography{akhanli-hennig-validation_SC_bibtex}


\end{document}